\documentclass[aps,twocolumn,prd,showpacs,nofootinbib]{revtex4}

\usepackage{amsmath}
\usepackage{graphicx}
\usepackage{amssymb}

\def\setC{\mathbb{C}}

\def\setR{\mathbb{R}}
\def\setC{\mathbb{C}}
\def\setN{\mathbb{N}}
\newcommand{\si}[1]{{\scriptscriptstyle{#1}}}

\newcommand{\dd}{\mathrm{d}}
\newcommand{\ee}{\mathrm{e}}
\newcommand{\DD}{\mathcal{D}}
\newcommand{\F}{H}
\newcommand{\matter}{\mathrm{mat}}

\newcommand{\rdim}{r}
\newcommand{\ie}{\textsl{i.e.~}}
\newcommand{\eg}{\textsl{e.g.~}}

\newcommand{\radim}{\rho}
\newcommand{\calF}{\mathcal{F}}
\newcommand{\calZ}{\mathcal{Z}}
\newcommand{\calT}{\mathcal{T}}
\newcommand{\calV}{\mathcal{V}_{_1}}
\newcommand{\calVV}{\mathcal{V}_{_2}}

\newcommand{\calP}{\mathcal{P}}
\newcommand{\calO}{\mathcal{O}}
\newcommand{\calS}{\mathcal{S}}
\newcommand{\sfx}{s_{_\mathrm{F}}}
\newcommand{\lfx}{l_{_\mathrm{F}}}
\newcommand{\mfx}{m_{_\mathrm{F}}}
\newcommand{\ffx}{f_{_\mathrm{F}}}
\newcommand{\Qfx}{Q_{_\mathrm{F}}}
\newcommand{\wfx}{w_{_\mathrm{F}}}
\newcommand{\zero}{{_{0}}}
\newcommand{\one}{{_{1}}}
\newcommand{\two}{{_{2}}}
\newcommand{\three}{{_{3}}}

\newcommand{\uh}{\mathrm{h}}
\newcommand{\uc}{\mathrm{c}}

\newcommand{\ug}{\mathrm{g}}
\newcommand{\ellg}{\ell_\ug}
\newcommand{\ellh}{\ell_\uh}
\newcommand{\sfnum}{\sqrt{\dfrac{2}{5}}}
\newcommand{\kappasix}{\kappa_{_6}}
\newcommand{\kappaw}{\kappa_{_w}}
\newcommand{\GReCO}{${\cal G}\setR\varepsilon\setC{\cal O}$}
\newcommand{\kappaf}{\kappa_{_f}}
\newcommand{\kappas}{\kappa_{_s}}

\newcommand{\Chi}{{\mathcal{X}}}

\newcommand{\gper}{\vartheta}
\newcommand{\Gper}{\Theta}
\newcommand{\gperR}{\gper_{r}}
\newcommand{\gperT}{\gper_{\theta}}
\newcommand{\GperR}{\Gper_{r}}
\newcommand{\GperT}{\Gper_{\theta}}
\newcommand{\Thetadim}{{\tilde{\varTheta}}}

\newcommand{\Upsiadim}{\tilde{\varUpsilon}}
\newcommand{\madim}{\tilde{M}}

\newcommand{\gTTi}{\frac{\ee^{-\gamma}}{r^2}}
\newcommand{\gTT}{r^2 \ee^{\gamma}}
\newcommand{\lngTTp}{\frac{1}{r} + \frac{\gamma'}{2}}

\newcommand{\Sigmadim}{\tilde{\varSigma}}

\newcommand{\Sigmadimdot}{\tilde{X}}
\newcommand{\Thetadimdot}{\tilde{W}}

\begin{document}

\title{Stability of six-dimensional hyperstring braneworlds}

\author{Christophe Ringeval}
\email{c.ringeval@imperial.ac.uk}
\affiliation{Blackett Laboratory, Imperial College, Prince Consort
Road, London SW7 2AZ, United Kingdom\\ D\'epartement de Physique
Th\'eorique, Universit\'e de Gen\`eve, 24 quai Ernest Ansermet, 1211
Gen\`eve 4, Switzerland}

\author{Patrick Peter} 
\email{peter@iap.fr} 
\affiliation{Institut d'Astrophysique de Paris, (\GReCO), UMR
7095-CNRS, 98bis boulevard Arago, 75014 Paris, France}

\author{Jean-Philippe Uzan} 
\email{uzan@iap.fr} 
\affiliation{Institut d'Astrophysique de Paris, 
(\GReCO), UMR 7095-CNRS, 98bis boulevard Arago, 75014 Paris, France}

\date{\today}

\begin{abstract}
We study a six-dimensional braneworld model with infinite warped
extra dimensions in the case where the four-dimensional brane is
described by a topological vortex of a U(1) symmetry-breaking Abelian
Higgs model in presence of a negative cosmological constant. A
detailed analysis of the microscopic parameters leading to a finite
volume space-time in the extra dimensions is numerically performed. As
previously shown, we find that a fine-tuning is required to avoid any
kind of singularity on the brane. We then discuss the stability of the
vortex by investigating the scalar part of the gauge-invariant
perturbations around this fine-tuned configuration. It is found that
the hyperstring forming Higgs and gauge fields, as well as the
background metric warp factors, cannot be perturbed at all, whereas
transverse modes can be considered stable. The warped space-time
structure that is imposed around the vortex thus appears severely
constrained and cannot generically support nonempty universe
models. The genericness of our conclusions is discussed; this will
shed some light on the possibility of describing our space-time as a
general six-dimensional warped braneworld.
\end{abstract}
\maketitle

\section{Introduction}\label{Sec:I}

Following the advent of string theory~\cite{Polchinski:1998rq} and its
implication that space may have more than the usual three dimensions
(in the Kaluza-Klein way) came the suggestion that the
extra dimensions could be much larger than previously expected, would
it be because of a smaller value of the Planck energy in the
bulk~\cite{Arkani-Hamed:1998rs,Arkani-Hamed:1998nn,
Antoniadis:1998ig}, or because of a large curvature in the (infinite)
extra dimensions~\cite{Randall:1999ee,Randall:1999vf}. A novel idea
came into play with the assumption that we live on a hypersurface, a
three-spatial dimensional ``brane'', embedded in a larger dimensional
warped space-time bulk~\cite{Rubakov:1983bb,Visser:1985qm}.

For any higher dimensional Universe model, it is essential to confine
gravity since gravitation is experimentally tested to be three
dimensional on many different scales, ranging from the
millimeter~\cite{Long:2002wn} to a few
Mega-parsecs~\cite{Uzan:2000mz,Allen:2000ih}. For a five-dimensional
anti-de Sitter bulk, gravity was shown to be localized on the
brane~\cite{Randall:1999ee, Randall:1999vf} and to lead to a viable
cosmological framework~\cite{Binetruy:1999ut, Csaki:1999jh,
Cline:1999ts, Binetruy:1999hy, Kraus:1999it, Shiromizu:1999wj,
Flanagan:1999cu, Maartens:1999hf, Rubakov:2001kp} provided the brane
and bulk cosmological constants are adjusted by hand. The situation is
not yet settled concerning the cosmological perturbations induced in
the brane~\cite{Riazuelo:2002mi}, although there are some indications
that such models should satisfy more stringent constraints than
previously expected~\cite{Ringeval:2003na}. In the case the brane is
modeled as a domain wall-like topological defect, this fine-tuning
transforms into a tuning of the underlying parameters (masses,
coupling constant and bulk cosmological
constant)~\cite{Ringeval:2001cq}.

Many mechanisms have been proposed to confine the other known
interactions and their associated particles:
scalar~\cite{Bajc:1999mh}, gauge bosons~\cite{Dvali:1996bg,
Dvali:1997xe, Dubovsky:2000am, Dvali:2000rx, Dimopoulos:2000ej,
Duff:2000jk, Oda:2001ux, Ghoroku:2001zu, Akhmedov:2001ny}, and
fermions~\cite{Neronov:2001qv}. In the latter case the mechanism
relies on a generalization of the cosmic string
case~\cite{Jackiw:1975fn, Ringeval:2001xd, Ringeval:2002qi}. The
fermions are trapped in the brane in the form of massless zero modes
and some can even become massive, although their mass spectrum is not
compatible~\cite{Ringeval:2003na} with the observed
one~\cite{PDBook}. It was also suggested~\cite{Dvali:2001qr}
that the electroweak Higgs field, and thus the origin of electroweak
symmetry breaking, could be understood from the existence of an
extra dimension in the form of a transverse gauge field
component. Most of these works were based on the simplifying
assumption of a reflection symmetry with respect to the brane,
although a more refined treatment, not assuming such a symmetry,
appears possible~\cite{Carter:2001nj, Battye:2001yn, Carter:2001af}.

Most of the relevant discussions on braneworld models have been
restricted to the case of one spatial extra dimension, as advocated
\eg in the framework of the eleven dimensional realization of
$M-$theory proposed by Ho\v{r}ava and
Witten~\cite{Horava:1996qa}. Moreover, the underlying brane model of
the Universe is often assumed to be infinitely thin in the transverse
direction, so that (i) the induced gravity stems essentially from
Darmois~--~Israel junction conditions~\cite{Binetruy:1999ut,
Csaki:1999jh, Cline:1999ts, Binetruy:1999hy, Kraus:1999it,
Shiromizu:1999wj, Flanagan:1999cu, Maartens:1999hf, Rubakov:2001kp}
and (ii) is mostly independent of the microstructure of the brane, if
any. No such general condition is available in the less restrictive
situation of more than one extra dimension that is the subject of the
present work.

To study braneworlds with more than one extra dimension, it is
necessary to specify the microstructure of the brane to fix a model,
taking into account in particular the possibly finite thickness of the
brane~\cite{Tinyakov:2001jt}. In particular, one needs to regularize
long range self-interaction forces (including gravity). In the case
the force derives from a potential (\eg in linearized gravity), the
self-interaction potential is well behaved outside the brane but will
be singular on the brane in the thin brane limit. One needs to
introduce a UV cutoff associated with the underlying
microstructure~\cite{Carter:2002tk}. The only case where such a
procedure can be avoided is that of hypermembranes that can be
satisfactorily treated without recourse to regularization.

Much work has been devoted to warped geometries in six
dimensions~\cite{Burgess:2001bn, Kogan:2001yr, Gogberashvili:2000hu,
Gogberashvili:2000yq, Gogberashvili:2001jm, Cohen:1999ia,
Kanti:2001vb}. When considering explicit realizations in terms of an
underlying topological defect forming field
model~\cite{Gregory:1999gv, Gregory:1997wk, Nihei:2000gb,
Gherghetta:2000qi}, it seems that the six-dimensional case represents
a limiting situation: for more than two extra dimensions, it was found
that it is not possible to confine gravity on a strictly local brane
although global topological defect configurations which allow gravity
confinement might exist~\cite{Gherghetta:2000jf, Roessl:2002rv}. At
least two questions arise: are the properties of gravity dependent on
the microscopic structure of the brane and is the chosen microscopic
model consistent with $M-$theory?  The second question has started to
be addressed in Ref.~\cite{Antunes:2002hn}, and it turns out that, for
many purposes, it is possible to consider defectlike realizations of
branes, as in the present article.

Before considering the possibility of
trapping~\cite{Giovannini:2002sb, Giovannini:2002jf} particle fields
in a hyperstring embedded in an anti-de Sitter six-dimensional bulk
space-time (adS$_{_6}$), it is necessary to determine the background
structure itself within a given field content, and decide whether it
is possible to localize gravity in the Universe thus obtained, thereby
generalizing the five-dimensional case. This article is devoted to
this task and accordingly models the brane by a vortex of a U(1)
Abelian gauge-Higgs model. By means of a numerical exploration of the
parameter space, we discuss different classes of solution exhibiting
an anti-de Sitter space-time at infinity. It appears that they are
generically associated with a conical or curvature singularity at the
brane location, except for some fine-tuning between the model
parameters. In the regular case, our approach agrees with the
numerical results of Ref.~\cite{Giovannini:2001hh}.

Assuming this fine-tuning, we then go on to analyze the stability of
the regular solution by studying the scalar modes of the gauge
invariant perturbations, as originally suggested in
Ref.~\cite{Giovannini:2002mk}. Restricting our attention to the lowest
angular momentum modes, we show that the only acceptable perturbations
of the nonvanishing background quantities, \ie the hyperstring
forming fields and metric warp factors, are the vanishing ones. The
conditions imposed on these perturbations to be acceptable being to be
initially finite with respect to the background value away from the
brane (a condition necessary in order to ensure that the bulk is close
to anti-de Sitter space), and to be well-behaved on the brane. We also
study the perturbations of quantities which are not involved in the
background configurations, as radial gauge field components and
nondiagonal metric perturbations. It is found that this subset of
perturbations is stable if the requirement of being bounded far from
the hyperstring is relaxed (a condition which may not be required
since there is not reference background fields for these
perturbations).

The article is organized as follows: after setting the field theoretic
framework both for the particles and gravity in Sec.~\ref{Sec:II}, we
construct the Nielsen-Olesen ansatz for a three-dimensional vortex
configuration, set and discuss the corresponding Euler-Lagrange field
equations in Sec.~\ref{Sec:III}. We then show how to handle the
boundary conditions in Sec.~\ref{Sec:IV} and solve numerically the
field equations in Sec.~\ref{Sec:V}, insisting in particular on the
numerous technical difficulties. This permits us to obtain the
parameter range over which gravity is localized and exempt of
singularity in the core. We then discuss some arguments leading to the
suggestion that such a defect realization of a brane in six dimensions
may be marginal in Sec.~\ref{Sec:VI}, and explicitly derive and
discuss its allowed perturbations in Sec.~\ref{sec:pert} before ending
by some concluding remarks.

\section{Vortex configuration in adS$_{_6}$}\label{Sec:II}

We consider the action for a complex scalar field $\Phi$ coupled to
gravity in a six-dimensional space-time
\begin{equation}
\label{eq:action} S=\int\left[\frac{1}{2\kappa_{_6}^2}(R-2\Lambda)
+{\cal L}_{\matter}\right]\sqrt{-g}\,\dd^6 x,
\end{equation}
where $g_{\si{AB}}$ is the six-dimensional metric with signature
$(+,-,-,-,-,-)$, $R$ its Ricci scalar, $\Lambda$ the six-dimensional
cosmological constant and $\kappa_{_6}^2\equiv32\pi^2G_{_6}/3$,
$G_{_6}$ being the six-dimensional gravity constant\footnote{In $D$
dimensions, we relate $\kappa_\si{D}^2$ to the $D$-dimensional
gravitational constant $G_\si{D}$ by
$\kappa_\si{D}^2=(D-2)\Omega^{[\si{D}-2]}G_\si{D}$, where
$\Omega^{[\si{D}-1]}=2\pi^{D/2}/\Gamma(D/2)$ is the surface of the
$(D-1)$-sphere.}. The matter Lagrangian reads
\begin{equation}\label{eq:lag}
{\cal L}_{\matter} =
\frac{1}{2}g^{\si{AB}}\DD_{\si{A}}\Phi\left(\DD_{\si{B}}\Phi\right)^*
-V(\Phi)-\frac{1}{4}\F_{\si{AB}}\F^{\si{AB}},
\end{equation}
in which capital Latin indices $A,B\ldots$ run from 0 to 5,
$\F_{\si{AB}}$ is the electromagneticlike tensor defined by
\begin{equation}\label{eq:fab}
\F_{\si{AB}}=\partial_{\si{A}} C_{\si{B}} -\partial_{\si{B}}
C_{\si{A}},
\end{equation}
where $C_{\si{B}}$ is the connection 1-form. The U(1) covariant
derivative $\DD_{\si{A}}$ is defined by
\begin{equation}\label{eq:d}
\DD_{\si{A}}\equiv\partial_{\si{A}} -iqC_{\si{A}},
\end{equation}
where $q$ is the charge. The potential of the scalar field $\Phi$ is
chosen to break the underlying U(1) symmetry and thereby allow for
topological vortex (cosmic stringlike) configurations,
\begin{equation}\label{eq:V}
V(\Phi)=\frac{\lambda}{8}\left(\vert\Phi\vert^2-\eta^2\right)^2,
\end{equation}
where $\lambda$ is a coupling constant and $\eta=\langle
|\Phi|\rangle$ is the magnitude of the scalar field vacuum expectation
values (VEV)\footnote{Note, that, because of the unusual number of
space-time dimensions, the fields have dimensions given by $[R]=M^2$,
$[\Phi]=M^{2}$, $[\Lambda]=M^2$, $[\lambda]=M^{-2}$, $[\eta]=M^{2}$,
$[\kappa_{_6}^2]=M^{-4}$, $[C_{\si{A}}]=M^2$, $[\F_{\si{AB}}]=M^3$ and
$[q]=M^{-1}$ ($M$ being a unit of mass). This can be further
generalized in the $D-$dimensional case by: $[\Phi]=M^{(D-2)/2}$,
$[\lambda]=M^{4-D}$, $[\eta]=M^{(D-2)/2}$, $[\kappa_{_D}^2]=M^{2-D}$,
$[C_{\si{A}}]=M^{D-2}$, $[\F_{\si{AB}}]=M^{D-3}$ and $[q]=M^{2-D/2}$,
$[R]$ and $[\Lambda]$ being unchanged.}.

Motivated by the brane picture, we choose the metric of the bulk
space-time to be of the warped static form
\begin{equation}\label{eq:metric}
\dd s^2=g_{\si{AB}} \dd x^{\si{A}} \dd x^{\si{B}} =
\ee^{\sigma(\rdim)}\eta_{\mu\nu} \dd x^\mu\dd x^\nu
-\dd\rdim^2-\rdim^2\ee^{\gamma(\rdim)}\dd\theta^2,
\end{equation}
where $\eta_{\mu\nu}$ is the four-dimensional Minkowski metric of
signature ($+,-,-,-$), and $(\rdim,\theta)$ the polar coordinates in
the extra dimensions. Greek indices $\mu,\nu\ldots$ run from 0 to 3
and describe the brane world sheet and we set $g_{\mu\nu} \equiv
\exp [{\sigma(\rdim)}] \eta_{\mu\nu}$. The action (\ref{eq:action})
with the ansatz (\ref{eq:metric}) will admit static solutions
depending only on $\rdim$ so that the general covariance along the
four-dimensional (physical) space-time is left unbroken.

The Nielsen-Olesen like~\cite{Nielsen:1987fy} ansatz for a generalized
vortex configuration is taken to be of the form~\cite{Peter:1992dw}
\begin{equation}\label{eq:ansatz}
\Phi= \varphi(\rdim)\ee^{i n\theta} = \eta f(\rdim)\ee^{i
n\theta},\qquad C_\theta= \displaystyle
\frac{1}{q}\left[n-Q(\rdim)\right]
\end{equation}
where $n$ is an integer, so that the only nonvanishing component of
the electromagnetic tensor is $\F_{\theta\rdim}=Q'/q$. With such an
ansatz, we shall now derive the relevant field equations and discuss
their solutions.

\section{Equations of motion}\label{Sec:III}

With the metric given by Eq.~(\ref{eq:metric}), the nonvanishing
Einstein tensor components reduce to
\begin{eqnarray}\label{eq:G}
G_{\mu\nu} &=&
\frac{1}{4}g_{\mu\nu}\bigg(6\sigma''+\frac{6}{\rdim}\sigma'
+6\sigma'^2+3\sigma'\gamma'\nonumber \\ & &
+2\gamma''+\gamma'^2+ \frac{4}{\rdim}\gamma'\bigg), \nonumber\\
G_{\rdim\rdim}&=&-\frac{1}{2}\sigma'\left(3\sigma'+\frac{4}{\rdim}+2\gamma'
\right), \nonumber\\
G_{\theta\theta}&=&-\frac{1}{2}\rdim^2\hbox{e}^\gamma\left(4\sigma''
+5\sigma'^2\right),
\end{eqnarray}
where a prime denotes differentiation with respect to
$\rdim$. Similarly, the matter stress-energy tensor,
\begin{equation}
\label{eq:tmunumatt}
T_{\si{AB}} \equiv 2\frac{\delta {\cal L}_{\matter}}{\delta
g^{\si{AB}}} - g_{\si{AB}} {\cal L}_{\matter},
\end{equation}
has nonvanishing components provided by the Nielsen-Olesen
ansatz~(\ref{eq:ansatz}) that are given by
\begin{eqnarray} \label{eq:tmunu}
T_{\mu\nu} &=&
g_{\mu\nu}\left[V+\frac{\eta^2}{2}\left(f'^2+\frac{Q^2f^2}{\rdim^2}
\hbox{e}^{-\gamma}\right)+\frac{1}{2}\frac{Q'^2}{q^2\rdim^2}\hbox{e}^{-\gamma}
\right],\nonumber\\ T_{\rdim\rdim} &=& -V
+\frac{\eta^2}{2}\left(f'^2-\frac{Q^2f^2}{\rdim^2}\hbox{e}^{-\gamma}\right)
+\frac{1}{2}\frac{Q'^2}{q^2\rdim^2}\hbox{e}^{-\gamma} ,\nonumber\\
T_{\theta\theta} &=& \rdim^2\hbox{e}^\gamma\left[ -V
-\frac{\eta^2}{2}\left(f'^2-\frac{Q^2f^2}{\rdim^2}\hbox{e}^{-\gamma}\right)
+\frac{1}{2}\frac{Q'^2}{q^2\rdim^2}\hbox{e}^{-\gamma} \right]
.\nonumber \\ & &
\end{eqnarray}
It follows that the six-dimensional Einstein equations, with our
conventions,
\begin{equation}
\label{eq:einstein}
G_{\si{AB}} + \Lambda g_{\si{AB}} + \kappa_{_6}^2 T_{\si{AB}} = 0,
\end{equation}
can be cast in the form
\begin{eqnarray}
\frac{3}{2}\ddot\sigma+\frac{3}{2}\dot\sigma^2+\frac{3}{2\radim}\dot\sigma
+\frac{3}{4}\dot\sigma\dot\gamma+\frac{1}{2}\ddot\gamma
+\frac{1}{4}\dot\gamma^2+\frac{1}{\radim}\dot\gamma=-\frac{\Lambda}{|\Lambda|}
\hfill \nonumber\\ -\alpha\left[\beta\left(f^2-1\right)^2+\dot f^2
+\frac{\hbox{e}^{-\gamma}}{\radim^2}\left(Q^2f^2+\frac{\dot
Q^2}{\varepsilon} \right)\right],\label{eq:einstein1}
\end{eqnarray}

\begin{eqnarray}
\frac{3}{2}\dot\sigma^2+\frac{2}{\radim}\dot\sigma
+\dot\sigma\dot\gamma&=&-\frac{\Lambda}{|\Lambda|}
-\alpha\Bigg[\beta\left(f^2-1\right)^2\nonumber \\ &-& \dot f^2
+\frac{\hbox{e}^{-\gamma}}{\radim^2}\left(Q^2f^2-\frac{\dot
Q^2}{\varepsilon} \right)\Bigg],\label{eq:einstein2}
\end{eqnarray}

\begin{eqnarray}
2\ddot\sigma+\frac{5}{2}\dot\sigma^2&=&-\frac{\Lambda}{|\Lambda|}
-\alpha\Bigg[\beta\left(f^2-1\right)^2\nonumber \\ &+&\dot f^2
-\frac{\hbox{e}^{-\gamma}}{\radim^2}\left(Q^2f^2+\frac{\dot
Q^2}{\varepsilon} \right)\Bigg],\label{eq:einstein3}
\end{eqnarray}
where we have introduced the dimensionless radial coordinate
\begin{equation}\label{eq:dimless1}
\radim\equiv\sqrt{\vert\Lambda\vert}\rdim,
\end{equation}
as well as the dimensionless parameters
\begin{equation}\label{eq:dimless2}
\alpha\equiv\frac{1}{2}\kappa_{_6}^2\eta^2,\qquad
\beta\equiv\frac{1}{4}\frac{\lambda\eta^2}{|\Lambda|},\qquad
\varepsilon\equiv\frac{q^2\eta^2}{|\Lambda|}.
\end{equation}
In Eqs.~(\ref{eq:einstein1}) to (\ref{eq:einstein3}), we have
introduced the convention that a dot refers to a differentiation with
respect to the dimensionless radial coordinate $\radim$.

The scalar field dynamics is given by the Klein-Gordon equation
\begin{eqnarray}\label{eq:KG}
\nabla_{\si{A}}\nabla^{\si{A}}\Phi&=&-\frac{\lambda}{2}
\left(\vert\Phi\vert^2-
\eta^2\right)\Phi +q^2C^2\Phi \\ \nonumber &
&+iqC^{\si{A}}\partial_{\si{A}}\Phi+iq\nabla_{\si{A}}
\left(C^{\si{A}}\Phi\right),
\end{eqnarray}
which takes the reduced form
\begin{equation}\label{eq:KGred}
\ddot f +\left(2\dot\sigma
+\frac{1}{2}\dot\gamma+\frac{1}{\radim}\right)\dot f =
\frac{Q^2}{\radim^2}f\hbox{e}^{-\gamma}+2\beta\left(f^2-1\right)f,
\end{equation}
while the Maxwell equations
\begin{equation}\label{eq:max}
\nabla_{\si{A}}
\F^{\si{AB}}=-q^2C^2\vert\Phi\vert^2+\frac{i}{2}q\left(
\Phi\partial^{\si{B}}\Phi^*-\Phi^*\partial^{\si{B}}\Phi\right),
\end{equation}
provide the single reduced equation for the only nonvanishing
component of the gauge vector field
\begin{equation}\label{eq:maxred}
\ddot Q + \left(2\dot\sigma-\frac{1}{2}\dot\gamma-
\frac{1}{\radim}\right)\dot Q = \varepsilon f^2 Q.
\end{equation}

The set of equations (\ref{eq:einstein1}, \ref{eq:einstein2},
\ref{eq:einstein3}, \ref{eq:KGred}, \ref{eq:maxred}) is a set of five
differential equations for 4 unknown functions ($\sigma$, $\gamma$,
$f$, $Q$).  Indeed, we have a redundant equation due the Bianchi
identities and one can check that the Higgs field equation
(\ref{eq:KGred}) is recovered from the constraint equation
(\ref{eq:einstein2}) provided $\dot{f} \ne 0$. This system can be
further simplified by remarking that Eq.~(\ref{eq:maxred}) can also be
written as

\begin{equation}
\frac{\dd}{\dd \radim} \left( \frac{\ee^{2\sigma}}{\sqrt{m}} \dot Q \right) =
\frac{\ee^{2\sigma}}{\sqrt{m}} \varepsilon f^2 Q,
\label{eq:QQdot1}
\end{equation}
where
\begin{equation}
\label{eq:mvdefs}
m \equiv \radim^2 v \equiv \radim^2 \ee^\gamma,
\end{equation}
so that
\begin{equation}
\frac{\dd}{\dd \radim} \left( \frac{\ee^{2\sigma}}{\sqrt{m}} Q \dot Q
\right) = \frac{\ee^{2\sigma}}{\sqrt{m}}\left( \varepsilon f^2
Q^2+\dot Q^2\right).
\label{eq:QQdot2}
\end{equation}
On the other hand, combining Eqs.~(\ref{eq:einstein1}) and
(\ref{eq:einstein3}), one also finds that
\begin{eqnarray}
\frac{\dd}{\dd \radim} \left[ \ee^{2\sigma}\sqrt{m} \left(
as+bl\right) \right] &=& -\ee^{2\sigma}\sqrt{m}\Bigl[ \left(a+b\right)
\calF \nonumber \\ 
& &+\frac{1}{2} \left(3b-a\right) \calV + 2 b\calVV \Bigr],
\nonumber \\
& & \label{eq:AAdot}
\end{eqnarray}
for any set of arbitrary constants $a$ and $b$, and where we have set
the new functions
\begin{equation}
\label{eq:lmdefs}
s \equiv \dot{\sigma}, \qquad l \equiv \dfrac{\dot{m}}{m} =
\dfrac{2}{\rho} + \dot{\gamma},
\end{equation}
as well as
\begin{eqnarray}
\label{eq:funcdefs} {\calF} &\equiv& \alpha\beta
\left(f^2-1\right)^2 +\frac{\Lambda}{|\Lambda|},\\ \calV &\equiv&
\frac{2\alpha}{\varepsilon}\frac{\dot{Q}^2}{m}, \qquad \calVV \equiv
{2\alpha} \frac{f^2 Q^2}{m}.\label{eq:funcdefs2}
\label{eq:defs}\end{eqnarray}
Using the relation (\ref{eq:AAdot}) with $a=-b$ leads to
\begin{equation}
\frac{\dd}{\dd \radim} \left[ \ee^{2\sigma}\sqrt{m} (s-l) \right] =
\frac{4 \alpha \ee^{2\sigma}}{\sqrt{m}}\left( \frac{\dot
Q^2}{\varepsilon} + f^2 Q^2\right),
\label{eq:QQdot3}
\end{equation}
and combining this result with Eq.~(\ref{eq:QQdot2}), one obtains
\begin{equation}
\frac{4\alpha}{\varepsilon} \dfrac{Q\dot Q}{m} = s - l +
c,\label{eq:QQdot}
\end{equation}
where $c$ is a remaining integration constant~\cite{Verbin:1998tc}. The
equations of motion (\ref{eq:einstein1}, \ref{eq:einstein2},
\ref{eq:einstein3}, \ref{eq:KGred}, \ref{eq:maxred}) end up being
equivalent to the following five first-order differential equations
\begin{eqnarray}
\label{eq:s}
\dot{s} & = & -\dfrac{5}{2} s^2 + \dfrac{2 \alpha}{\varepsilon}
\dfrac{Q w}{m} s - \calF + \frac{1}{2} \calV - \dfrac{c}{2}s , \\
\label{eq:m}
\dot{m}  & = & s m - \dfrac{4 \alpha}{\varepsilon} Q w + c m,\\
\label{eq:f}
\alpha \dot f^2 & = & \dfrac{5}{2} s^2 - \dfrac{4 \alpha}{\varepsilon}
\dfrac{Q w}{m} s + \calF + \dfrac{\calVV}{2} - \dfrac{\calV}{2} + c s,
\\
\label{eq:Q}
\dot Q & = & w, \\
\label{eq:w}
\dot w & = & \varepsilon Q f^2 -\dfrac{2\alpha}{\varepsilon} \dfrac{Q
w^2}{m} -\dfrac{3}{2} s w + \dfrac{c}{2}  w.
\end{eqnarray}

After discussing the behavior of these fields far from the vortex, \ie
far in the bulk, and on the brane itself in the following section, we
shall solve numerically the field equations in order to determine the
structure of the space-time and defect system.

\section{Asymptotic behaviors}\label{Sec:IV}

By definition of the topological defect like configuration, we require
that the Higgs field vanishes on the membrane itself, \ie $\Phi = 0$
for $\radim=0$, while it recovers its VEV, $\eta$, in the bulk. These
requirements translate into the following boundary conditions for the
function $f$:
\begin{equation}
\label{eq:limhiggs}
f(0)=0,\qquad \lim_{\radim\to +\infty}f = 1.
\end{equation}
The corresponding boundary conditions for the 1-form connection are
given by
\begin{equation}
\label{eq:limgauge}
 Q(0)=n,\qquad \dot Q(0)=w(0)=0, \qquad\lim_{\radim\to +\infty}Q = 0.
\end{equation}

In order to avoid any curvature singularity on the string, the Ricci
scalar stemming from Eq.~(\ref{eq:action}), namely
\begin{equation}
\label{eq:ricci}
R=-|\Lambda| \left(\ddot{\gamma} + 4 \ddot{\sigma} + \frac{1}{2}
\dot{\gamma}^2 + 5 \dot{\sigma}^2 + 2 \dot{\gamma} \dot{\sigma} +
\frac{2 \dot{\gamma}}{\radim} + \frac{4 \dot{\sigma}}{\radim}\right),
\end{equation}
has to be finite at $\radim=0$ as the vortex is assumed to represent
our physical four-dimensional space. As a result, the warp functions
$\dot{\gamma}(0)$ and $\dot{\sigma}(0)$ have to vanish in the string
core,
\begin{equation}
\label{eq:warp0}
\dot{\gamma}(0)=\dot{\sigma}(0)=0,
\end{equation}
and the warp function $l$ therefore scales near the string like
\begin{equation}
\label{eq:l0}
l(\radim) \underset{0}{\sim} \frac{2}{\radim} \quad \Rightarrow \quad
m \underset{0}{\propto} \radim^2.
\end{equation}
Note that we impose both functions to vanish on the string, and not
merely the combination $\dot{\gamma}+2\dot{\sigma}$ entering
Eq.~(\ref{eq:ricci}); this arises from the requirement that \emph{all}
geometrical quantities, \eg $R^\si{A}_{\ \si{B}} R^\si{B}_{\ \si{A}}$
in which $\dot{\gamma}$ and $\dot{\sigma}$ enter with different
coefficients, must be finite. A coordinate transformation along the
brane allows to choose $\sigma(0)=0$, while $\gamma(0)$ and $v_\zero$,
defined by
\begin{equation}
\label{eq:v0}
v_\zero\equiv\ee^{\gamma(0)},
\end{equation}
are determined by the boundary conditions at infinity. Note that
$v_\zero$ cannot be absorbed in a rescaling of the radial
coordinate. To see this, it suffices to introduce a new coordinate,
$\tilde r$ say, such that $r^2 \ee^{\gamma(0)}=\tilde r^2$, which is
equivalent to defining $\gamma(\tilde r) = \gamma(r)-\gamma(0)$. This
would induce a shift in the other warp function $\sigma$, shift that
can however be taken care of by a rescaling of the vortex internal
coordinates. This is not all though, because the last metric element
$g_{rr}=-1$ gets modified into $g_{\tilde r \tilde r} = g_{rr}
\ee^{\gamma(0)} = - \ee^{\gamma(0)}$. All the derivatives with respect
to this new radial variable also acquire this numerical factor. In the
Einstein tensor, given the symmetries of the vortex, this seems
harmless as $G_{\si{AB}}$ is also simply rescaled. However, the
stress-energy tensor (\ref{eq:tmunu}) is not so simply rescaled as it
also involves non derivative terms (the Higgs field potential $V$ and
the gauge-Higgs coupling). On the other hand, $v_\zero$ can be
absorbed by a rescaling of the angular coordinate $\tilde{\theta} =
\sqrt{v_\zero} \theta$. In that case, the angular part of the metric
(\ref{eq:metric}) appears to be cylindrical in the hyperstring core,
with however a missing angle
\begin{equation}
\label{eq:anglecore}
\Delta \tilde \theta = 2 \pi \left( 1-\sqrt{v_\zero} \right).
\end{equation}
The space-time geometry obtained for $v_\zero \ne 1$ exhibits a conical
singularity in the vortex core whose physical interpretation is the
existence of an additional $\delta$-like energy-momentum distribution
(a Goto-Nambu hyperstring) lying at the center of the
configuration. This interpretation remains valid provided $0 \le
v_\zero \le 1$, the other cases will be discussed in the next
section. At this point, it is interesting to note that contrary to
what is assumed in Ref.~\cite{Giovannini:2001hh}, the value of $v_\zero$ in our
approach is completely determined as soon as the other boundary
conditions are imposed and ends up being a function of the model
parameters only. Setting $v_\zero=1$ afterward, to obtain a regular
geometry in the hyperstring core, will allow us to recover the
fine-tuning relation obtained in Ref.~\cite{Giovannini:2001hh}.

In the following, we derive analytical approximations of the fields at
infinity and in the hyperstring core associated with an anti-de Sitter
space-time at infinity. The influence of the model parameters on these
solutions is discussed.

\subsection{Far from the string}
\label{sect:far}

Asymptotically, the anti-de Sitter space-time is recovered provided
\begin{equation}
\label{eq:sinf}
\lim_{\radim \to +\infty} \dot{s} =\lim_{\radim \to +\infty}
\dot{l} = 0.
\end{equation}
Denoting by an index `{\small F}' (standing for ``fixed'') the value
of the fields at infinity, it follows from Eqs.~(\ref{eq:limhiggs}),
(\ref{eq:limgauge}) and (\ref{eq:sinf}) that the adS$_{_6}$ solution
is a fixed point for the set of Eqs. (\ref{eq:s}~--~\ref{eq:w}) with
$\ffx=1$, $\wfx=\Qfx=0$ for the Higgs and gauge fields, and with the
equations
\begin{eqnarray}
\label{eq:sfix}
-\dfrac{5}{2} \sfx^2 + \frac{c}{2} \sfx - \frac{\Lambda}{|\Lambda|} &
 = & 0,\\
\label{eq:mfix}
\mfx \sfx + c \mfx & = & 0, \\
\label{eq:ffix}
\dfrac{5}{2} \sfx^2 + 2 c \sfx + \frac{\Lambda}{|\Lambda|} & = & 0,
\end{eqnarray}
for the warp factors. The equations (\ref{eq:sfix}) and
(\ref{eq:ffix}) can only be simultaneously satisfied for $c=0$, since
$\sfx=0$ would lead to $\Lambda=0$. As a result, Eq.~(\ref{eq:mfix})
requires that $\mfx=0$ and the asymptotic warp factors reduce to
\begin{equation}
\label{eq:warpF}
\lfx^2 = \sfx^2 = -\frac{2}{5} \frac{\Lambda}{|\Lambda|}.
\end{equation}
The anti-de Sitter solution is obtained for $\Lambda <0$ so that
$\lfx=\sfx=-\sqrt{2/5}$ and we can now verify that $\calV$ and
$\calVV$ effectively vanish. Indeed, from Eq.~(\ref{eq:m}), the
dominant behavior of $m$ at infinity is given by
\begin{equation}
\label{eq:mF}
m \sim m_\infty \ee^{-\sqrt{2/5} \radim},
\end{equation}
while Eqs.~(\ref{eq:Q}) and (\ref{eq:w}) admit asymptotically the
decaying solutions
\begin{eqnarray}
\label{eq:gaugeF}
Q \sim Q_\infty \ee^{-\ellg \radim}, \qquad w \sim -\ellg Q,
\end{eqnarray}
with
\begin{equation}
\label{eq:defellg}
\ellg = \frac{3}{4} \sqrt{\frac{2}{5}}\left(\sqrt{\frac{40}{9}
\varepsilon +1} - 1 \right).
\end{equation}
{}From Eq.~(\ref{eq:defs}), the gauge functions $\calV$ and $\calVV$
vanish at infinity provided
\begin{equation}
\label{eq:epsilonmin}
2 \ellg > \sqrt{\frac{2}{5}} \quad \Leftrightarrow \quad \varepsilon >
\frac{2}{5}.
\end{equation}
This is the first restriction on the available parameter space.

{}From Eq.~(\ref{eq:KGred}), the decaying branch of the Higgs field at
infinity is given by
\begin{equation}
\label{eq:higgsinf}
h \sim h_\infty \ee^{ -\ellh \radim} +
\dfrac{Q_\infty^2}{4 m_\infty} \dfrac{\ee^{-\left( 2
\ellg - \sqrt{2/5} \right)\radim}}{\beta - \left(\ellg -
\dfrac{1}{2}\sfnum\right)\left( \ellg +
\dfrac{3}{4}\sfnum \right)},
\end{equation}
where we have defined
\begin{equation}
h \equiv 1-f,
\end{equation}
and $\ellh$ reads
\begin{equation}
\label{eq:defellh}
\ellh=\dfrac{1}{2} \sqrt{\dfrac{5}{2}} \left(\sqrt{\dfrac{32}{5}\beta
  + 1} -1 \right).
\end{equation}
{}From Eq.~(\ref{eq:higgsinf}), it appears that the behavior of the
Higgs field at infinity can be driven by the gauge field, provided
$\ellh > 2 \ellg - \sqrt{2/5}$. In this case, one can check that $h$
remains positive definite ensuring that the Higgs field approaches its
VEV from below and thus can support a topological defect
configuration. Indeed, would $h$ be negative asymptotically, the
condition $f(0)=0$ could only be realized if the sign of the
derivative $\dot{f}$ changes at some intermediate point. {}From
Eq.~(\ref{eq:KGred}), it is clear that at the point $\dot{f}=0$, the
right-hand side is positive for $f>1$. As a result $\ddot{f}$ can only
be positive and the profile of the Higgs field would remain always
convex and greater than its VEV.

\begin{figure*}
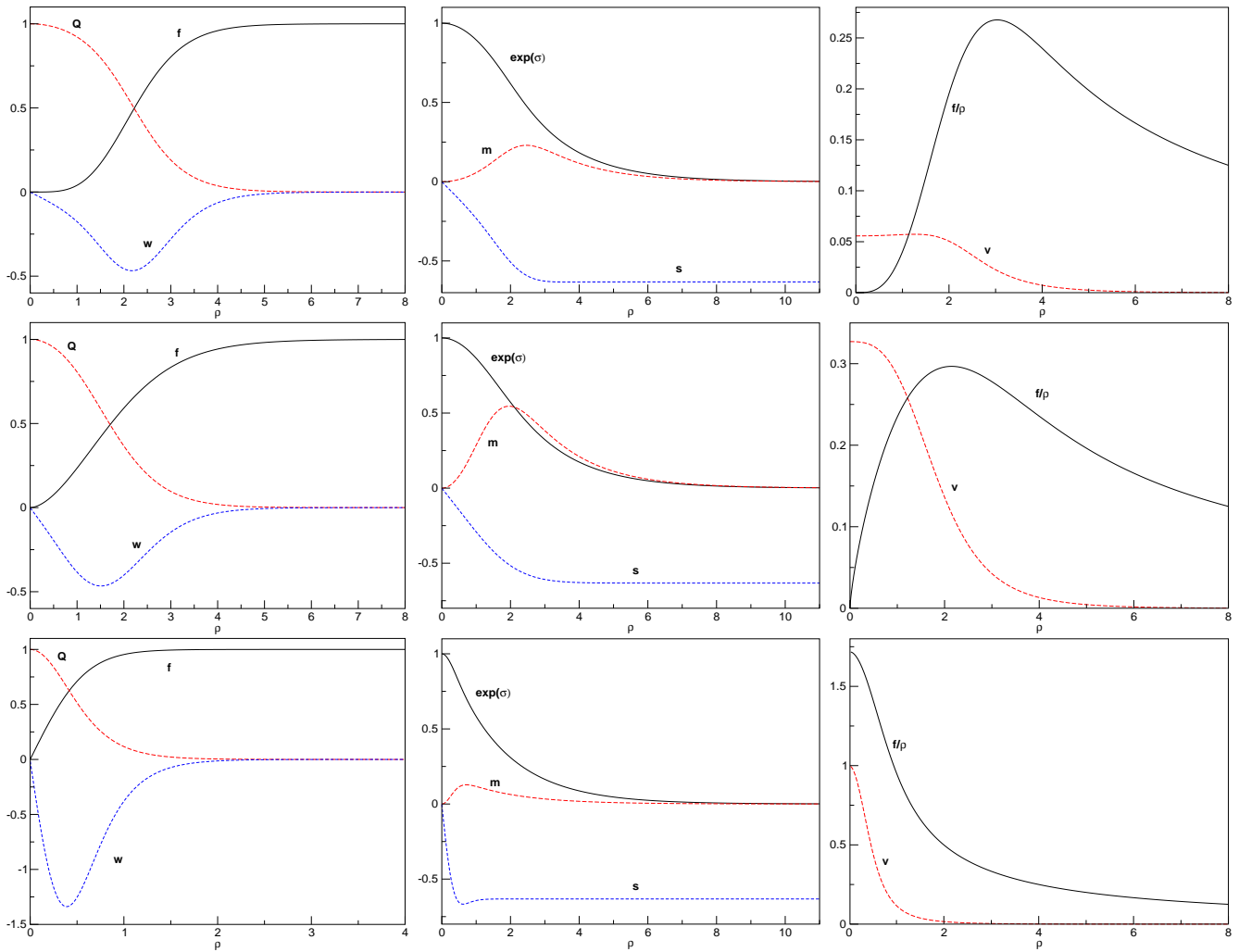

\begin{center}
\includegraphics[width=5.8cm,height=4.5cm]{fig1a.eps}
\includegraphics[width=5.8cm,height=4.5cm]{fig1b.eps}
\includegraphics[width=5.8cm,height=4.5cm]{fig1c.eps}\\
\includegraphics[width=5.8cm,height=4.5cm]{fig1d.eps}
\includegraphics[width=5.8cm,height=4.5cm]{fig1e.eps}
\includegraphics[width=5.8cm,height=4.5cm]{fig1f.eps}\\
\includegraphics[width=5.8cm,height=4.5cm]{fig1g.eps}
\includegraphics[width=5.8cm,height=4.5cm]{fig1h.eps}
\includegraphics[width=5.8cm,height=4.5cm]{fig1i.eps}
\caption{Typical field solutions associated with an anti-de Sitter
space-time at infinity in the regimes where $v_\zero < 1/4$, $ 1/4<
v_\zero < 1$ and $v_\zero \sim 1$, obtained for
$(\alpha,\varepsilon,\beta)$ equals to $(1.00,5.00,1.50)$,
$(2.20,5.00,0.80)$ and $(1.29,14.83,6.39)$ respectively. In each case,
the Higgs and gauge fields are plotted on the left picture, the warp
factors are plotted in the middle one, while the right plot represents
the behavior of $f/\radim$ and $v \equiv m/\radim^2$. Note the
behavior of the warp factor $s$ (middle plots) for the parameters
living above or under the plane (\ref{eq:paraplane}) (see also
Fig.~\ref{fig:isosurfs} and Fig.~\ref{fig:regsurf}).}
\label{fig:solgrav}
\end{center}
\end{figure*}

Similarly, the asymptotic expression for the warp factor can be
obtained from Eq.~(\ref{eq:s}). In terms of the new function
\begin{equation}
u \equiv s + \sfnum ,
\end{equation}
one gets for the decaying branch, up to some fine-tuning (see
Sect.~\ref{Sec:VI})
\begin{equation}
\label{eq:uinf}
u \sim \dfrac{4 \alpha \beta}{2 \ellh + \sqrt{10}} h_\infty \ee^{-2
  \ellh \radim} - \dfrac{\alpha \ellg}{2
  \epsilon}\dfrac{Q_\infty^2}{m_\infty} \ee^{-\left(2 \ellg -
  \sqrt{2/5} \right)\radim}.
\end{equation}
If $2\ellh < 2 \ellg - \sqrt{2/5}$, the convergence of the warp
factors toward the anti-de Sitter solution is driven by the Higgs
field and $u$ remains positive definite since $h_\infty$ is positive
in that case [see Eq.~(\ref{eq:higgsinf})]. On the other hand, if
$2\ellh > 2 \ellg - \sqrt{2/5}$, the metric factor $s$ behaves
asymptotically like the gauge field and $u$ remains definite
negative. As the result, the surface $2\ellh = 2 \ellg - \sqrt{2/5}$,
\ie from Eqs.~(\ref{eq:defellg}) and (\ref{eq:defellh}),
\begin{equation}
\label{eq:paraplane}
\beta = -\dfrac{1}{10} + \dfrac{1}{4} \varepsilon,
\end{equation}
separates the parameter space $(\alpha,\varepsilon,\beta)$ in two
regions where the warp factor $s$ approaches its anti-de Sitter value
from above or below, respectively (see Fig.~\ref{fig:solgrav}).

\subsection{Near the string}
\label{sect:near}

The field behaviors near the string, to leading order in $\radim$, can
be extracted from the equations of motion (\ref{eq:s}) to
(\ref{eq:w}).  We assume the truncated Taylor expansions
\begin{equation}
w \sim w_\one \radim^{\kappaw}, \qquad Q \sim n +
\frac{w_\one}{1+\kappaw} \radim^{1+\kappaw},
\label{eq:Qw0}
\end{equation}
for the gauge fields, and
\begin{equation}
\label{eq:f1}
f \sim f_\one \radim^{\kappaf},
\end{equation}
for the Higgs field with $\kappaw \ge 1$ and $\kappaf \ge 1$ in order
that $\dot{w}$ and $\dot{f}$ remains finite in $\radim=0$. Similarly,
the warp factor $s$ can be expanded around $\radim=0$ as
\begin{equation}
s \sim s_\one \radim^{\kappas}, \label{eq:s0}
\end{equation}
with $\kappas \ge 1$ in order for the Ricci scalar to remains finite
in the core [see Eq.~(\ref{eq:ricci})]. Setting $c=0$ (see
Sect.~\ref{sect:far}) in Eq.~(\ref{eq:w}), using Eqs.~(\ref{eq:l0}),
(\ref{eq:v0}), (\ref{eq:Qw0}) and (\ref{eq:f1}), yields $\kappaw=1$
and
\begin{equation}
v_\zero = -\frac{2\alpha}{\varepsilon} n w_\one.
\label{eq:v0w1}
\end{equation}
{}From Eq.~(\ref{eq:s}) and making use of Eqs.~(\ref{eq:funcdefs}) and
(\ref{eq:funcdefs2}), one gets
\begin{equation}
\label{eq:s1}
\kappas=1, \qquad s_\one=\frac{1}{2}\left(1 - \alpha \beta +
\frac{\alpha}{\varepsilon} \frac{w_\one^2}{v_\zero} \right).
\end{equation}
The Higgs field behavior is given by Eq.~(\ref{eq:f}) and reads
\begin{equation}
\alpha \kappaf^2 f_\one^2 \radim^{2 \kappaf-2}=2 s_\one +\left(\alpha
\beta -1\right) - \frac{\alpha}{\varepsilon} \frac{w_\one^2}{v_\zero} + 
\alpha \frac{f_\one^2 n^2}{v_\zero} \radim^{2 \kappaf -2}.
\end{equation}
With $s_\one$ given by the expression (\ref{eq:s1}), this implies
\begin{equation}
\label{eq:powerf}
\kappaf=\frac{n}{\sqrt{v_\zero}}.
\end{equation}
As a result, the behavior of the Higgs field around the string core is
only determined by the asymptotic solutions at infinity through
$v_\zero$ (and $w_\one$), as announced above, and the requirement
$\kappaf \ge 1$ only provides the constraint
\begin{equation}
\label{eq:v0cond}
v_\zero \le n^2.
\end{equation}

{}From a purely classical point of view, only $\dot{f}$ has to be well
defined at $\radim=0$. However, in the general situation, there always
exists an integer $p \in \setN$ such that $v_{\zero} > n^2/p$, in
which case all derivatives of the Higgs field, $f^{(k)}$, with $k\geq
p$ are divergent in the core since the $p^{\mathrm{th}}$ derivative
already is. This is so unless the bound is saturated, \ie the equality
$v_\zero=n^2/p$ is strictly satisfied so that $\kappaf=p$ and then all
derivatives of order larger than $p$ will strictly vanish. If the
bound is not saturated however, the metric at the core exhibits the
conical singularity on which the field reacts by introducing the
aforementioned divergences. Note that, as discussed above, the conical
singularity interpretation demands $v_\zero \le 1$, while for a vortex
with $n>1$, Eq.~(\ref{eq:v0cond}) allows $v_\zero$ to be larger than
one. This unexpected regularity comes from the scaling properties of
the field equations. Indeed, in Eqs.~(\ref{eq:s}) through
(\ref{eq:w}), the warp function $m$ (and thus $v$) appears only
through the ratio $Q^2/m$ and $w^2/m$. As a result, a solution for a
given winding number $n$ and $v_\zero$, is also solution of a winding
number $\tilde{n}=pn$ and $\tilde{v}_\zero=p^2 v_\zero$. This scaling
permits to solve the equations for a reduced set of parameters and yet
obtain the complete spectrum of solutions. On the physical side, the
kind of singularity appearing in the core for $1 < v_\zero \le n^2$
may be interpreted as an supercritical Goto-Nambu hyperstring lying in
$\radim=0$. On the other hand, the behavior of the fields leading to
$v_\zero> n^2$ requires divergences of $\dot{f}$ and the stress tensor
no longer remains finite in the vortex core [see
Eq.~(\ref{eq:tmunu})]. In that case, the conditions $\kappaw \ge 1$,
$\kappaf \ge 1$ and Eqs.~(\ref{eq:s1}) are no longer valid and the
regularity requirements in Eq.~(\ref{eq:warp0}) are no longer
satisfied such that a curvature singularity appears in $\radim=0$.

The previous analysis makes clear that the value of $v_\zero$ encodes
the regularity of the matter fields and the geometry in the string
core. By requiring the space-time to be of anti-de Sitter kind at
infinity, we have shown that $v_\zero$ is a function of the model
parameters only, as long as the boundary conditions
(\ref{eq:limhiggs}), (\ref{eq:limgauge}) and (\ref{eq:warpF}) can be
satisfied. In the following, we numerically recover the field behavior
expected from the asymptotic analysis, in particular the regular
configurations in the core require a fine-tuning in the model
parameters such as $v_\zero=1$.

\section{Numerical approach and problems}
\label{Sec:V}

Several technical difficulties appear in the numerical integration of
the equations of motion (\ref{eq:einstein1}) to (\ref{eq:maxred}). The
first issue comes from the requirement of an anti-de Sitter space-time
far from the core. Indeed, as mentioned in Sect.~\ref{sect:far}, the
metric factors as well as the Higgs and gauge fields admit growing
modes at infinity which correspond to infinite-volume space-time in
the extra dimensions (see Sect.~\ref{Sec:VI} and
Appendix~\ref{App:6dLambda}). As the result, any direct numerical
integration starting from the core toward the outer regions will
necessary jump, due to the finite numerical accuracy, onto these
growing solutions. This numerical instability can be overcome by
performing a backward integration starting from a finite cutoff
distance far the hyperstring toward the core. However, one has to pay
attention to choose a convenient set of functions, as $s$ and $m$
rather than $\sigma$ and $\gamma$ to remove any explicit $1/\radim$
dependencies in the equations of motion. Indeed, the effect of the
$1/\radim$ term in Eqs.~(\ref{eq:einstein1}) to (\ref{eq:maxred}) is
to add flow turning points where the signs of the field derivatives
change. As a result, growing behaviors would appear from these turning
points whatever the initial conditions and the direction of
integration. There is no such flow inversions in the closed system of
Eqs.~(\ref{eq:s}) to (\ref{eq:w}), and the exponential growth can be
suppressed by integrating from the anti-de Sitter fixed point at
infinity toward the hyperstring core. However, one has still to face
the following difficulty. As pointed in Eq.~(\ref{eq:uinf}), even the
decaying branch of $s$ at infinity admits two exponential decaying
modes: one varying as $\exp{(-2 \ellh \radim)}$ and the other as
$\exp{[-(2 \ellg -\sqrt{2/5}) \radim]}$. As a result, a backward
direct integration is numerically unstable with respect to one of
these modes. In the same way that a forward direct integration would
jump onto the growing mode, such a backward numerical method would
tend to select by numerical finite accuracy the strongest decaying
exponential. For instance, if $2 \ellh > 2 \ellg -\sqrt{2/5}$, a
backward method would be preferentially sensitive to the $\exp{(-2
\ellh \radim)}$ behavior (by moving toward the lower values of
$\radim$, this mode blows up faster than the other), which is not the
one in which we are interested. It is interesting to note that the
numerical instability of the backward method disappears for $2
\ellh=2\ellg -\sqrt{2/5}$ where the two decaying modes are identified,
which means that this method can only be efficient on a surface of the
parameter space $(\alpha,\varepsilon,\beta)$ whose equation is given
by Eq.~(\ref{eq:paraplane}). Note that a similar numerical instability
occurs in Eq.~(\ref{eq:KGred}) between the two corresponding decaying
modes of Eq.~(\ref{eq:higgsinf}).

In order to overcome these difficulties, we have chosen to use a
finite difference numerical method instead of a direct method. In
particular, the equations of motion (\ref{eq:einstein1}) to
(\ref{eq:maxred}) deriving from the action (\ref{eq:action}), we have
implemented a successive over-relaxation
method~\cite{Adler:1983zh}. By discretizing the radial coordinate
$\radim$, the action (\ref{eq:action}) can be expressed as a finite
sum over $\radim_i$ (the integer $i$ indexing the discrete values
taken by $\radim$ over the radial grid) whose differentiation with
respect to the fields evaluated at the discrete points leads to a
system of finite difference equations corresponding to
Eqs.~(\ref{eq:einstein1}), (\ref{eq:einstein3}), (\ref{eq:KGred}) and
(\ref{eq:maxred}). {}From a initial guess of the field and metric
profiles along the radial grid, the values taken by the fields at
$\radim_i$ are corrected by a Newton's method to reduce the error with
respect to the true solution. In this approach, the boundary
conditions are part of the finite difference equations since they
appear as the conditions satisfied by the fields at the first and last
point of the $\radim_i$ grid (see Appendix A in
Ref.~\cite{Ringeval:2004ju} for a relaxation method applied to a
similar action). This procedure is stable provided the initial guessed
profiles are not too far from the true solutions. The iterative
corrections can been stopped when the discrete action remains
stationary at the machine precision. We have also checked that the
numerical solutions obtained in this way satisfy the constraint
equation (\ref{eq:einstein2}).

\begin{figure*}
\begin{center}
\includegraphics[width=16cm]{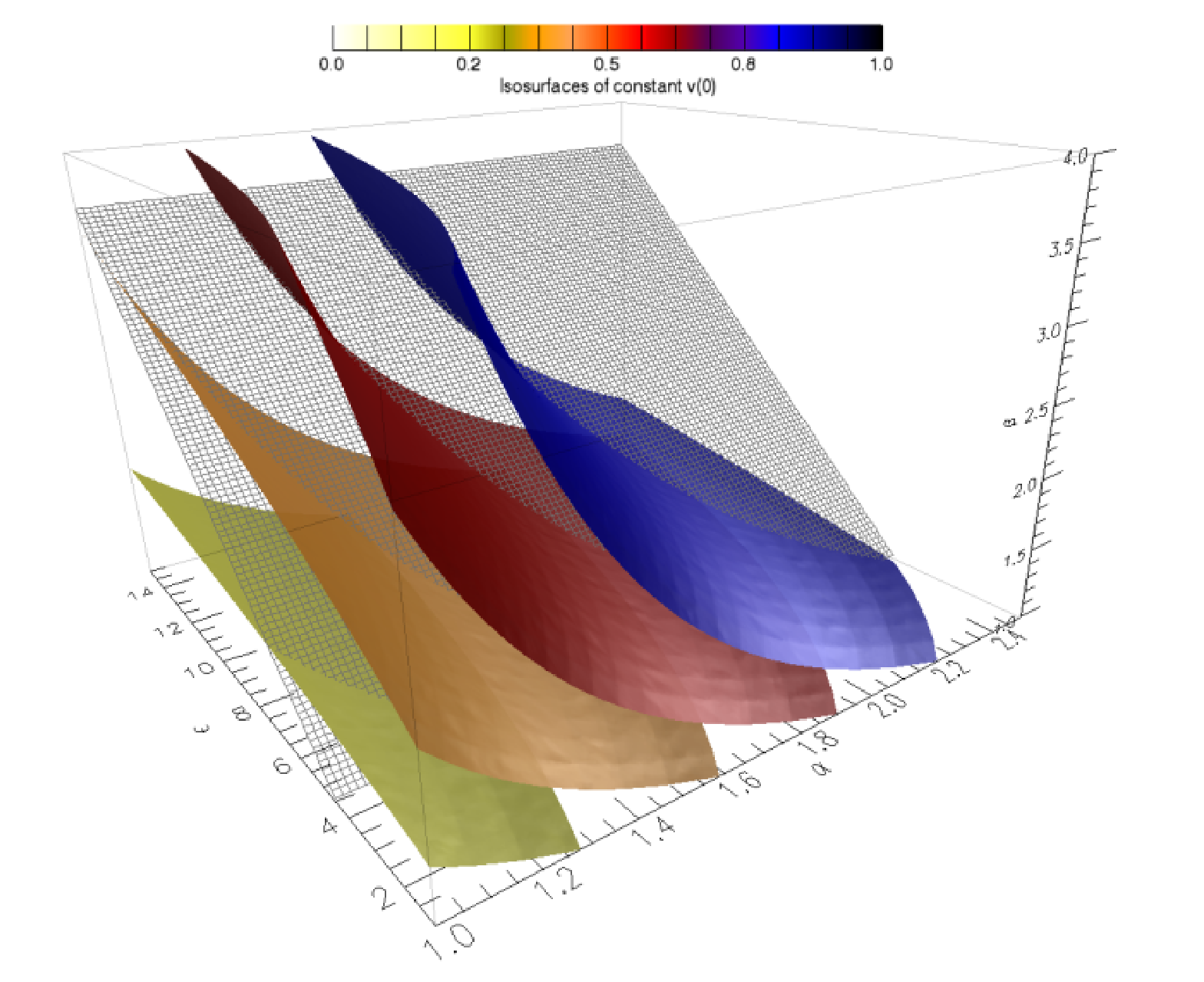}
\caption{Isosurfaces of constant $v_\zero$ in the parameter space
$(\alpha,\varepsilon,\beta)$ associated with an anti-de Sitter
space-time at infinity. {}From the left to the right, the four surfaces
correspond to $v_\zero=0.25$, $v_\zero=0.5$, $v_\zero=0.75$ and
$v_\zero=1$, respectively. The wired mesh represents the plane
$\beta=-1/10 + \varepsilon/4$ (see Sect.~\ref{sect:far}) which
separates the parameter space in two regions. For $\beta > -1/10 +
\varepsilon/4$ the gauge field drives the warp factors toward their
anti-de Sitter value and $s$ has a global minimum at a finite distance
to the core whereas, in the other case, the Higgs field dominates at
infinity and $s$ always decreases toward its asymptotic value (see
Fig.~\ref{fig:solgrav}). On the left side of the $v_\zero=1$
isosurface the hyperstring exhibits a conical singularity in
$\radim=0$ whereas on the right side there is a curvature singularity
(see Fig.~\ref{fig:sing}). Only the surface $v_\zero=1$ ends up be
associated with a regular configuration on the brane (see
Fig.~\ref{fig:regsurf})}
\label{fig:isosurfs}
\end{center}
\end{figure*}

\begin{figure*}
\begin{center}
\includegraphics[width=15cm]{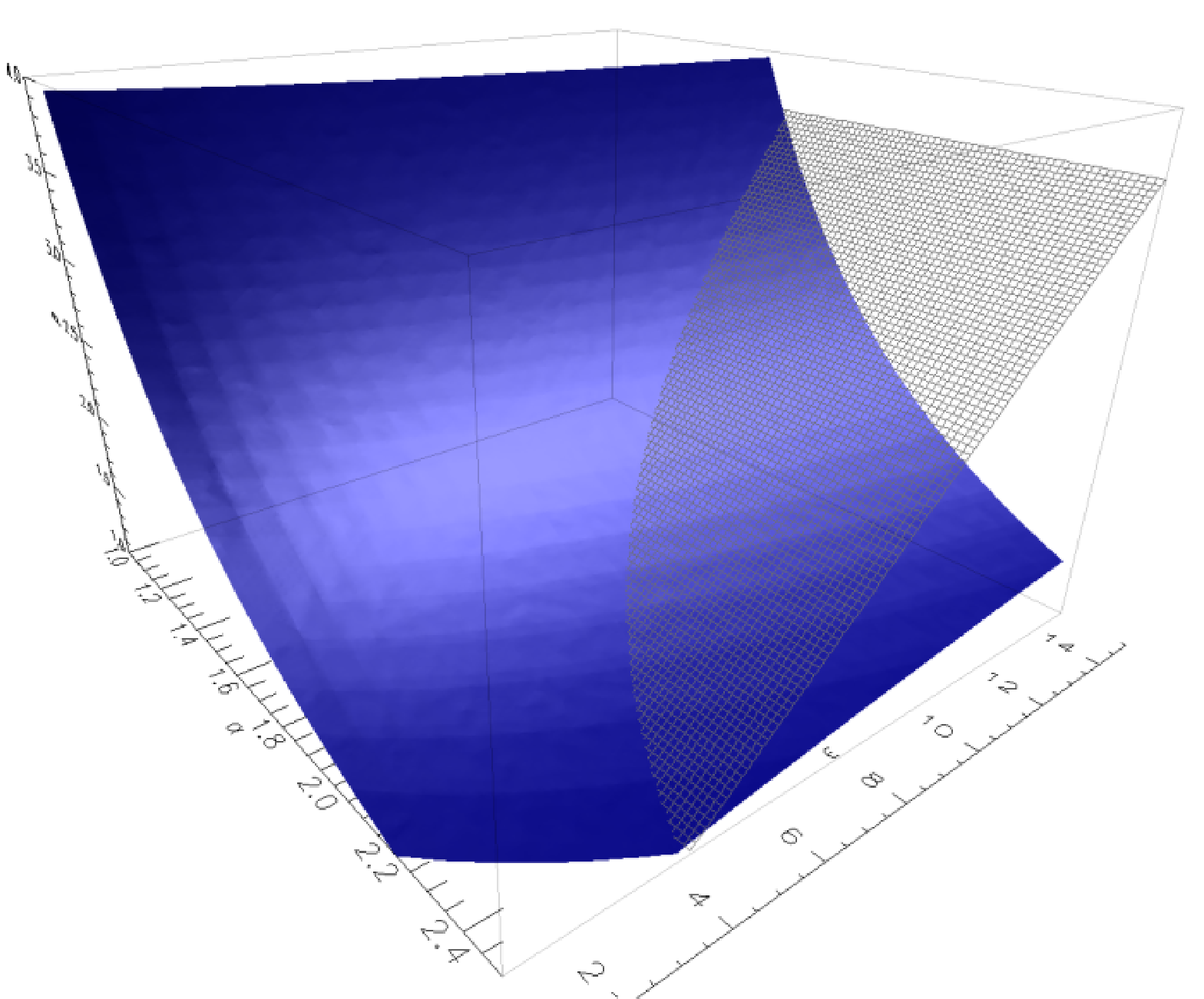}
\caption{The fine-tuning surface $v_\zero=1$ in the parameter space
$(\alpha,\varepsilon,\beta)$ associated with an anti-de Sitter
space-time at infinity and a regular geometry in the hyperstring core
(for the $n=1$ vortex). The wired mesh is the surface
$\beta=-1/10+\varepsilon/4$ (see Sect.~\ref{sect:far}).}
\label{fig:regsurf}
\end{center}
\end{figure*}

In order to probe the behavior of the solutions according to the model
parameters, we have, in a first time, numerically allowed regular and
conical solutions in the hyperstring core by requiring the boundary
conditions (\ref{eq:limhiggs}), (\ref{eq:limgauge}), (\ref{eq:warp0}),
(\ref{eq:l0}) and (\ref{eq:warpF}) to be satisfied. Along the lines
drawn in the previous paragraph, we have used the relaxation method to
compute the solutions of the field equations in the case of a unit
winding vortex $n=1$ and for $25^3$ values of the parameters
$(\alpha,\varepsilon,\beta)$. As expected from the asymptotic analysis
(see Sect.~\ref{Sec:IV}) the hyperstring generically develops a
conical singularity with $v_\zero \ne 1$ (see
Fig.~\ref{fig:solgrav}). The regular solutions $v_\zero=1$ are
obtained only for the parameters lying on the surface plotted in
Fig.~\ref{fig:regsurf} which identifies to the fine-tuning surface
previously found in Ref.~\cite{Giovannini:2001hh}.

\begin{figure}
\begin{center}
\includegraphics[width=8.5cm]{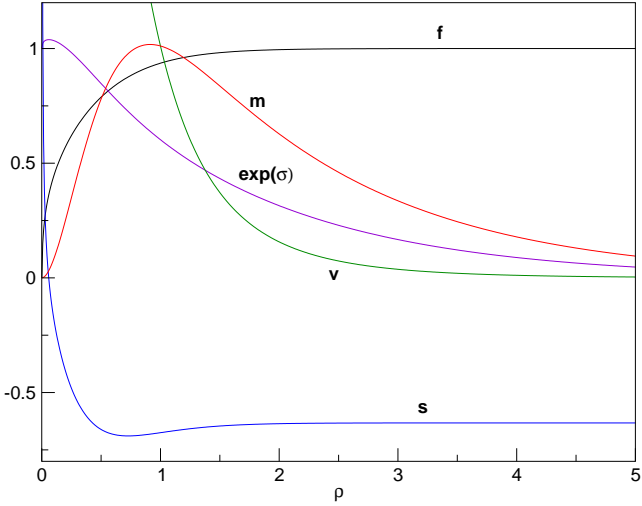}
\caption{Typical behavior of the Higgs field and metric factors for
  the model parameters living on the right-hand side of the
  $v_\zero=1$ isosurface, \ie for large value of $\alpha$ (see
  Fig.~\ref{fig:regsurf}). The derivatives of the Higgs field and the
  metric coefficient diverge in $\radim=0$ leading to a curvature
  singularity [see Eq.~(\ref{eq:ricci})].}
\label{fig:sing}
\end{center}
\end{figure}

However, under the previous boundary conditions, the relaxation
procedure failed to converge in the regions of the parameter space
which could have been associated with values of $v_\zero>1$ (see
Fig.~\ref{fig:isosurfs}). For these regions, we have thus weakened the
boundary conditions in order that Eqs.~(\ref{eq:limhiggs}),
(\ref{eq:limgauge}) and (\ref{eq:warpF}) are still satisfied but with
\begin{equation}
\sigma(0)=0, \qquad m(0)=0,
\end{equation}
instead of Eqs.~(\ref{eq:warp0}) and (\ref{eq:l0}). In that case, the
method converges and the numerical solutions exhibit a divergence in
the Higgs field derivative while the derivative of the warp factors do
no longer vanish in $\radim=0$ (see Fig.~\ref{fig:sing}). {}From
Eq.~(\ref{eq:ricci}), these solutions correspond to a curvature
singularity in the hyperstring core and are not physical.

\section{Fine-tuning and stability}\label{Sec:VI}

The solutions we have obtained for the fields surrounding a branelike
vortex requires a fine-tuning of the underlying microscopic parameters
to be free of singularity in the string core and of finite volume in
the extra dimensions. These conditions are indeed the minimal
requirements for an acceptable warped braneworld model in
six-dimensions. In the following we discuss qualitatively why such
fine-tuning is expected as well as the differences appearing with
respect to the domain wall model in a five-dimensional anti-de Sitter
case~\cite{Ringeval:2001cq}. Our analysis is analogous to the one
developed in Ref.~\cite{Gregory:1999gv, Gregory:1997wk} in the case of
global vortex and leads to similar conclusions.

The asymptotic form of Eq.~(\ref{eq:s}) reads
\begin{equation}
\dot s+\frac{5}{4} s^2 =\frac{1}{2},
\label{s6}
\end{equation}
for which one obtains the general solution in the form
\begin{equation}
s = \sqrt{\frac{2}{5}} \times \frac{A_{_6} \ee^{c_{_6} \radim} -
\ee^{-c_{_6} \radim}}{{A_{_6} \ee^{c_{_6} \radim} + \ee^{-c_{_6} \radim}}},
\label{sol6}
\end{equation}
where $c_{_6}=\sqrt{5/8}$, and $A_{_6}$ is an arbitrary constant, to
be matched with the vortex interior solution. In the asymptotic
analysis of Sect.~\ref{Sec:IV}, this constant has been set to zero,
but clearly, if $A_{_6}\not= 0$, one has
\begin{equation}
\lim_{\radim\to +\infty} s(\radim) = \sqrt{\frac{2}{5}},
\end{equation}
leading to an exponentially divergent warp factor for the
metric~(\ref{eq:metric}). 

Only the particular value $A_{_6}=0$ can smoothly join the interior
metric to a six-dimensional anti-de Sitter asymptotic space-time. In
five dimensions, this choice happens to be imposed by the Einstein
constraint~\cite{Ringeval:2001cq}, but in the case at hand, because of
the extra degree of freedom provided by the other function $\gamma$,
this is an explicit choice and not a mandatory consequence of the
Einstein equations. In other words, the solution satisfying
$\lim_{\radim\to\infty} s =-\sqrt{2/5}$ is also a point in the phase
space from which all trajectories diverge. This is related to the fact
that in the limit where the field contribution in the stress-energy
tensor is negligible with respect to the bulk cosmological constant,
Eqs.~(\ref{eq:einstein1}) and (\ref{eq:einstein3}) are two dynamical
equations for the warp functions $\sigma$ and $\gamma$, with
Eq.~(\ref{eq:einstein2}) being a constraint. The two first order (in
$\dot\sigma$ and $\dot\gamma$) dynamical equations require two
constants of integration, of which the constraint fixes only one
(together with the requirement of an anti-de Sitter asymptotic
space-time). The solution with arbitrary (nonvanishing) $A_{_6}$ is
thus a valid solution, contrary to the five-dimensional case. However,
any nonzero value of $A_{_6}$ does not correspond to a solution with
gravity localized on the vortex, being exponentially far from the
adS$_6$ case (see also the Appendix~\ref{App:6dLambda}).  One is thus
led to conclude that the fine-tuning required in the 6D case is worse
than in 5D since the physically relevant solution is a set of measure
zero in the full set of solutions. This drives us to ask whether such
a solution, although mathematically acceptable, can be reached by any
dynamical evolution. In fact, as we show in the following section, no
acceptable perturbation mode can be found for the regular vortex,
except for some special modes, the transverse ones, which have no
equivalent at zeroth order. In other words, the regular 6D vortex
configurations cannot be subject to perturbations in their background
fields and in particular, cannot depart infinitesimally from adS$_6$.

\section{gauge-invariant perturbations}\label{sec:pert}

As vector and tensor perturbations have been investigated
elsewhere~\cite{Giovannini:2002sb, Giovannini:2002jf}, we shall
concentrate on the scalar part~\cite{Giovannini:2002mk} of the
perturbations. Here and in what follows, the Scalar-Vector-Tensor
decomposition is understood to be with respect to the four-dimensional
vortex internal coordinates. Note that most of previous works were
concentrating on zero modes, which were found to be ``not
normalizable''. In what follows, we consider massive modes and
attention will be paid to the physical interpretation of the results.

\subsection{gauge-invariant variables}

In order to conclude on the stability (or physical relevance as we
shall see) of the configuration we obtained, it is necessary to
perturb this background solution in a gauge-invariant way. The first
order perturbation of the metric (\ref{eq:metric}), when restricting
attention to the scalar perturbations, reads
\begin{equation}
\label{eq:pertmetric}
\begin{aligned}
\dd s^2 &= \ee^{\sigma(r)} \left[ \eta_{\mu\nu} \left( 1+\psi \right)
 + \partial_\mu \partial _\nu E \right] \dd x^\mu \dd x^\nu -
(1+\xi)\dd r^2 \\ & - 2 \zeta \dd r\dd \theta
-r^2\ee^{\gamma(r)}\left(1+\omega\right) \dd \theta^2 \\ &
-2 \left( \partial _\mu B \dd r +\partial _\nu C \dd\theta \right)
\dd x^\mu ,
\end{aligned}
\end{equation}
where the scalar functions $\psi$, $E$, $\xi$, $\zeta$, $\omega$, $B$ and
$C$ depend on all the coordinates $(x^\alpha ,r,\theta)$ and are
assumed to be small.

A gauge transformation $x^{\si{A}} \to \tilde x^{\si{A}} = x^{\si{A}}
+ \epsilon^{\si{A}}$, with $\epsilon^\mu = \partial^\mu \epsilon$
(scalar transformations only) implies three gauge degrees of freedom,
so we are left with four unknown functions to determine. The scalar
functions transform under a gauge transformation as
\begin{equation}
\begin{aligned}
\tilde{\psi} & = \psi - \sigma' \epsilon_r, \qquad
\tilde E = E + 2 \ee^{-\sigma} \epsilon , \qquad
\tilde\xi = \xi - 2 \epsilon_r',\\
 \tilde \zeta &= \zeta -
\frac{1}{2} \left[\epsilon_\theta' +\partial_\theta
\epsilon_r - \left(\displaystyle{\frac{2}{r}}+\gamma'\right)
\epsilon_\theta \right],\\
 \tilde\omega &= \omega
- \frac{2\ee^{-\gamma}}{r^2} \partial_\theta\epsilon_\theta -
\left(\frac{2}{r}+\gamma'\right)\epsilon_r,\\
\tilde B &= B -  \frac{1}{2} \left( \epsilon_r + \epsilon'-\sigma'
\epsilon \right), \qquad
\tilde C = C - \displaystyle \frac{1}{2} \left( \epsilon_\theta +
\partial_\theta \epsilon \right).
\end{aligned}
\end{equation}
{}From these relations, we can derive the four gauge-invariant
variables
\begin{equation}
\begin{aligned}
 \Psi & \equiv \psi - \displaystyle \frac{1}{2} \sigma'
    \left(4B + \ee^\sigma E' \right), \quad
 \Xi  \equiv \xi
    -\partial_r\left(4B + \ee^\sigma E' \right),\\
 \Upsilon
    & \equiv\zeta - \displaystyle \frac{1}{4} \partial_r \left(4C +
    \ee^{\sigma} \partial_\theta E\right) -\displaystyle \frac{1}{4}
    \partial_\theta \left(4B + \ee^{\sigma} E'\right) \\
&  + \displaystyle\frac{1}{4} \left(\frac{2}{r} + \gamma'\right) \left(4C
    + \ee^{\sigma} \partial_\theta E \right),\\
 \Omega & \equiv \omega - \displaystyle \frac{\ee^{-\gamma}}{r^2}
    \partial_\theta \left(4C + \ee^\sigma \partial_\theta E \right) \\
 &     - \displaystyle \frac{1}{2} \left(\frac{2}{r} +
    \gamma' \right) \left(4B + \ee^\sigma \partial_r E
    \right).
\end{aligned}
\end{equation}

As in the usual cosmological case~\cite{Mukhanov:1990me}, these variables are
identical to the original variables once the choice of longitudinal
gauge ($E=B=C=0$) is made. The transformation leading to this gauge,
starting from an arbitrary gauge transformation, reads
\begin{equation}
\begin{aligned}
\epsilon &= \displaystyle -\frac{1}{2} \ee^\sigma E, \quad
\epsilon_r = \displaystyle 2B+\frac{1}{2} \ee^\sigma E', \quad
\epsilon_\theta = \displaystyle 2C+\frac{1}{2} \ee^\sigma
\partial_\theta E, 
\end{aligned}
\end{equation}
and is unique. This gauge choice, which we shall for now on adopt, is
thus complete for metric perturbations, but also for the matter
ones. Indeed, in our model (\ref{eq:lag}), the matter perturbations
concern only the hyperstring forming scalar field $\Phi$ and its
associated gauge field $C_{\si{A}}$. Note that in our framework, the
location of the brane is given by the zeroes of the Higgs field and
thus directly taken into account in its perturbations. Since we are
only interested in scalar perturbations, the perturbed fields can be
expanded as
\begin{equation}
\label{eq:perthiggs}
\begin{aligned}
\delta \Phi & = \chi(r,\theta) \, \ee^{in\theta}, \qquad \delta C_{\si
A} = \left(\partial_\mu \gper,\gperR,\gperT \right),
\end{aligned}
\end{equation}
where we have extracted the background winding phase $\ee^{in \theta}$
in the scalar field perturbations. Note that, for consistency, all the
perturbations have to be invariant by a complete rotation around the
hyperstring, and thus can be decomposed in discrete angular momentum
modes around the string. Under the gauge transformation $x^{\si A} \to
\tilde x^{\si A} = x^{\si A} + \epsilon^{\si A}$ these perturbations
transform to
\begin{equation}
\begin{aligned}
\tilde{\chi} & = \displaystyle \chi - \varepsilon_r \varphi' - i n
\frac{\ee^{-\gamma}}{r^2} \varepsilon_\theta \varphi, \qquad
\tilde{\gper}  =
\displaystyle \gper -\frac{\ee^{-\gamma}}{r^2} C_\theta
\varepsilon_\theta, \\
\tilde{\gperR} & = \displaystyle \gperR -
\frac{\ee^{-\gamma}}{r^2} C_\theta \varepsilon_{\theta}' + 2
\frac{\ee^{-\gamma}}{r^2} C_\theta \left(\frac{1}{r} +
\frac{\gamma'}{2} \right) \varepsilon_\theta,\\
\tilde{\gperT} & =
\displaystyle \gperT - C_{\theta}' \varepsilon_r -
\frac{\ee^{-\gamma}}{r^2} C_\theta \partial_\theta \varepsilon_\theta,
\end{aligned}
\end{equation}
and are therefore not invariant. Similarly to the metric tensor
decomposition, we then define the gauge-invariant quantities through
the relations
\begin{align}
\Chi &\equiv\displaystyle \chi - \frac{1}{2} \varphi'\left( 4 B +
\ee^\sigma E' \right) -\frac{i n \ee^{-\gamma}}{2 r^2} \varphi
\left(4C + \ee^\sigma \partial_\theta E \right),
\end{align}
for the Higgs perturbations and
\begin{align}
\Gper &\equiv \displaystyle \gper - \frac{1}{2}
\frac{\ee^{-\gamma}}{r^2} C_\theta \left(4C+\ee^{\sigma}
\partial_\theta E \right),\\ \GperR & \equiv \displaystyle \gperR -
\frac{1}{2} \frac{\ee^{-\gamma}}{r^2} C_\theta \left[4C'+
\left(\ee^{\sigma} \partial_\theta E\right)' \right] \nonumber \\ & +
\displaystyle \frac{\ee^{-\gamma}}{r^2} C_\theta \left(\frac{1}{r} +
\frac{\gamma'}{2} \right) \left(4C + \ee^{\sigma} \partial_\theta E
\right),\\ \GperT &\equiv \displaystyle \gperT - \frac{1}{2} C_\theta'
\left(4B+\ee^{\sigma} E' \right) \nonumber \\ & - \displaystyle
\frac{1}{2} \frac{\ee^{-\gamma}}{r^2} C_\theta \left(4 \partial_\theta
C + \ee^{\sigma} \partial^2_\theta E \right),
\end{align}
for the gauge field perturbations. They also match with the original
variables in the longitudinal gauge ($E=B=C=0$).

Note that since we are interested in perturbation theory, we have to
keep in mind that all the perturbed physical quantities involved at
some initial time have to be close to the background solution. This
implies in particular that we must impose on the physically meaningful
perturbations to be initially bounded: of all the possible solutions
of the perturbation equations which we discuss below, we shall retain
only those for which neither long- nor short-distance, divergence
appear. This, as it turns out, is extremely restrictive.

\subsection{Perturbed Einstein equations}

The Einstein equations, perturbed at first order, stem from
Eq.~(\ref{eq:einstein}). The perturbed metric tensor $\delta
g_\si{AB}$ is explicitly written in Eq.~(\ref{eq:pertmetric}) and
allows, by means of Eq.~(\ref{eq:perthiggs}), the determination of the
scalar part of the perturbed Einstein and stress-energy tensors. They
are derived in Appendix~\ref{App:dvp}, and Eq.~(\ref{eq:einstein})
leads, in terms of the gauge-invariant variables, to the following
equations of motion
\begin{widetext}
\begin{equation}
\label{eq:einstein_munu}
\begin{aligned}
\left(\partial_\mu \partial_\nu - \eta_{\mu \nu} \Box \right) \left(
\frac{\Xi + \Omega}{2} + \Psi \right) & +  \frac{1}{2} \ee^\sigma
\eta_{\mu \nu} \Bigg\{3 \Psi'' + 3 \gTTi \partial^2_\theta \Psi +
\Omega'' - 2 \gTTi \partial_\theta \Upsilon' + \gTTi \partial_\theta^2
\Xi + 3 \left(2 \sigma' + \lngTTp \right) \Psi'  \\ & - 
\left. \left(\frac{3}{2} \sigma' + \lngTTp \right) \Xi' +
\left[\frac{3}{2} \sigma' + 2 \left(\lngTTp\right) \right] \Omega' - 3
\gTTi \sigma' \partial_\theta \Upsilon \right.  \\ & + 
\left[3\sigma'' + 3\sigma'^2 + 3\sigma' \left(\lngTTp\right) +
2\left(\lngTTp\right)' + 2\left(\lngTTp\right)^2 \right] \left(\Psi -
\Xi \right) \Bigg\}  \\ & +  \kappasix^2 \ee^{\sigma}
\eta_{\mu \nu} \Bigg\{-\gTTi \frac{Q'}{q} \left(\GperT' -
\partial_\theta \GperR \right) +\varphi' \varSigma' + \gTTi
\varphi Q^2 \varSigma  + \gTTi \varphi Q \partial_\theta
\varDelta + \dfrac{\dd V}{\dd \varphi} \varSigma   \\
& -  \frac{1}{2}\left( \gTTi \frac{Q'^2}{q^2} + \varphi'^2 \right)
\Xi - \frac{1}{2} \gTTi \left(\frac{Q'^2}{q^2} + \varphi^2 Q^2 \right)
\Omega  \\ & +  \left[\frac{1}{2} \gTTi \left(
\frac{Q'^2}{q^2} +\varphi^2 Q^2 \right) + \frac{1}{2} \varphi'^2 +
V(\varphi) \right] \Psi - \gTTi q \varphi^2 Q \GperT \Bigg\} =
-\Lambda \ee^{\sigma} \Psi \eta_{\mu\nu},
\end{aligned}
\end{equation}
for the $(\mu,\nu)$ part of Eq.~(\ref{eq:einstein}). The $(\mu,r)$ and
$(\mu,\theta)$ components lead to the following equations,
respectively,
\begin{eqnarray}
\label{eq:einstein_mur}
3 \Psi' + \Omega' - \gTTi \partial_\theta \Upsilon - \left(\frac{3}{2}
\sigma' + \lngTTp \right) \Xi + \left(-\frac{1}{2} \sigma' + \lngTTp
\right) \Omega   + 2 \kappasix^2 \left\{- \gTTi \frac{Q'}{q}
\left(\GperT - \partial_\theta \Gper \right) + \varphi' \varSigma'
\right\} = 0,
\end{eqnarray}
and
\begin{eqnarray}
\label{eq:einstein_mutheta}
\partial_\theta\left(3 \Psi + \Xi \right) - \Upsilon' - \left(\sigma'
+ \lngTTp \right) \Upsilon + 2 \kappasix^2 \left\{ \frac{Q'}{q}
\left(\GperR - \Gper' \right) + \varphi Q \varDelta -q \varphi^2 Q
\Gper \right\} = 0.
\end{eqnarray}
The purely bulk components of the first order perturbation of
Eq.~(\ref{eq:einstein}) read
\begin{eqnarray}
\label{eq:einstein_rr}
\frac{1}{2} \ee^{-\sigma} \Box\left(3 \Psi + \Omega\right) & - & 2
\gTTi \partial_\theta^2 \Psi + 2 \gTTi \sigma' \partial_\theta
\Upsilon - \left[3 \sigma' + 2 \left(\lngTTp \right) \right] \Psi' -
\sigma' \Omega' \nonumber \\ & + & \kappasix^2 \Bigg\{-\gTTi
\frac{Q'}{q}\left(\GperT' - \partial_\theta \GperR \right) + 
\varphi' \varSigma' - \gTTi \varphi Q^2 \varSigma - \gTTi \varphi Q
\partial_\theta \varDelta - \dfrac{\dd V}{\dd \varphi} \varSigma
 \nonumber \\ & - & \left[\frac{1}{2} \gTTi \varphi^2 Q^2 +
V(\varphi) \right] \Xi - \frac{1}{2} \gTTi \left(\frac{Q'^2}{q^2} -
\varphi^2 Q^2 \right) \Omega + \gTTi q \varphi^2 Q \GperT \Bigg\} -
\Lambda \Xi = 0,
\end{eqnarray}
for the $(r,r)$ part,
\begin{eqnarray}
\label{eq:einstein_thetatheta}
\frac{1}{2} \gTT \ee^{-\sigma} \Box \left(3 \Psi + \Xi \right) & - & 2
\gTT \Psi'' + \gTT \sigma' \left(\Xi' - 5 \Psi' \right) + \frac{1}{2}
\gTT \left(4 \sigma'' + 5 \sigma'^2 \right) \left( \Xi - \Omega
\right) \nonumber \\ & + & \kappasix^2 \Bigg\{-\frac{Q'}{q}
\left(\GperT' - \partial_\theta \GperR \right) -\gTT \varphi'
\varSigma' + \varphi Q^2 \varSigma + \varphi Q \partial_\theta \varDelta
 - \gTT \dfrac{\dd V}{\dd \varphi} \varSigma
 \nonumber \\ & + & \frac{1}{2}
\left(\gTT \varphi'^2 - \frac{Q'^2}{q^2} \right) \Xi - \gTT \left[
\frac{1}{2} \varphi'^2 + V(\varphi) \right] \Omega - q \varphi^2 Q
\GperT \Bigg\} - \gTT \Lambda \Omega = 0,
\end{eqnarray}
for the $(\theta,\theta)$ component, while the mixed one $(r,\theta)$
leads to the equation
\begin{eqnarray}
\label{eq:einstein_rtheta}
{}-\frac{1}{2} \ee^{-\sigma} \Box \Upsilon & + & 2 \partial_\theta
\Psi' -\sigma' \partial_\theta \Xi + \left[\sigma' - 2
\left(\lngTTp\right) \right] \partial_\theta \Psi -\frac{1}{2} \left(4
\sigma'' + 5 \sigma'^2 \right) \Upsilon \nonumber \\ & + &
\kappasix^2 \Bigg\{\varphi Q \varDelta' - \varphi'Q \varDelta
+ \varphi' \partial_\theta \varSigma  + \left[\frac{1}{2} \gTTi\left(
\frac{Q'^2}{q^2} - \varphi^2 Q^2\right) - \frac{1}{2} \varphi'^2
-V(\varphi) \right] \Upsilon \nonumber \\ & - & q \varphi^2 Q \GperR
\Bigg\} - \Lambda \Upsilon = 0,
\end{eqnarray}
\end{widetext}
with ``$\Box$'' standing for the flat four-dimensional d'Alembertian, \ie
\begin{equation}
\Box = \eta^{\mu \nu} \partial_\mu \partial_\nu = \partial _t^2 -
\nabla^2,\label{nabla}
\end{equation}
and where the perturbed Higgs field has been decomposed as
\begin{equation}
\label{eq:complex_chi}
\Chi = \varSigma + i \varDelta,
\end{equation}
\ie into its real and imaginary parts.

The perturbation (\ref{eq:complex_chi}), although true in general,
does not make the most of the U(1) invariance of the theory
(\ref{eq:lag}). Indeed, with the definition (\ref{eq:d}) for the
covariant derivative, the full theory is unchanged under the change
$\Phi\to\Phi^\mathrm{(N)} = \ee^{i\alpha(x_{\si{A}})} \Phi$ provided
the gauge vector field is simultaneously modified into $C_{\si{A}} \to
C_{\si{A}}^\mathrm{(N)} = C_{\si{A}} -(1/q) \partial_{\si{A}}
\alpha$. For an infinitesimal U(1) gauge transformation, this leads to
the transformation $\varSigma^\mathrm{(N)}=\varSigma$ and
$\varDelta^\mathrm{(N)}=\varDelta + \alpha \varphi (r)$, so that
choosing $\alpha = -\varDelta/\varphi$ allows to restrict attention to
purely real perturbations of the scalar field perturbation $\Chi$; we
shall accordingly call this choice the real gauge. Indeed, as we shall
see explicitly, letting $\varDelta$ arbitrary leads to equations
involving only the U(1) gauge-invariant degrees of freedom, namely
$\Theta-\varDelta/(q\varphi)$, $\Theta_\theta -\partial_\theta
\varDelta/(q\varphi)$ and $\Theta_r-(1/q) (\varDelta/\varphi)'$, which
merely expresses the fact that by going to the real gauge, one can get
rid of $\varDelta$.

\subsection{Perturbed Maxwell equations}

By means of Eq.~(\ref{eq:perthiggs}), we can also derive the perturbed
Maxwell equations stemming from Eq.~(\ref{eq:max}), at first order in
the fields. Since the Einstein equations impose to the stress-energy
tensor to be conserved, the perturbed Maxwell equations are certainly
already included in Eqs.~(\ref{eq:einstein_munu}) to
(\ref{eq:einstein_rtheta}). Nevertheless, they mainly involve the
matter fields and may help to decouple the whole system. The $(\mu)$
component of the perturbed Faraday tensor gives
\begin{widetext}
\begin{eqnarray}
\label{eq:pertmax_mu}
\GperR' & - & \Gper'' + \left(\sigma' + \lngTTp \right) \left(\GperR -
\Gper' \right) + \gTTi \partial_\theta \left(\GperT - \partial_\theta
\Gper \right) - q \varphi \varDelta + q^2 \varphi^2 \Gper =
0,
\end{eqnarray}
while the $(r)$ and $(\theta)$ bulk parts lead to
\begin{eqnarray}
\label{eq:pertmax_r}
\ee^{-\sigma} \Box \left(\GperR - \Gper' \right) & + & \gTTi
\partial_\theta\left(\GperT' - \partial_\theta \GperR \right) - \gTTi
\frac{Q'}{q} \partial_{\theta} \left[ 2 \Psi - \frac{1}{2} \left(\Xi +
\Omega \right) \right] + \gTTi q \varphi^2 Q \Upsilon - q \varphi \varDelta'
 + q \varphi' \varDelta
\nonumber \\ & + & q^2 \varphi^2 \GperR = 0,
\end{eqnarray}
and
\begin{eqnarray}
\label{eq:pertmax_theta}
\ee^{-\sigma} \Box \left(\GperT - \partial_\theta \Gper \right) & - &
\left(\GperT'' -  \partial_\theta \GperR' \right) - \left[2 \sigma' -
\left(\lngTTp \right) \right] \left( \GperT' -\partial_\theta \GperR
\right) + \frac{Q'}{q} \left[2 \Psi' -\frac{1}{2} \left( \Xi'+ \Omega'
\right) \right] \nonumber \\ & - & \left\{ \frac{Q''}{q} + \left[2
\sigma'- \left(\lngTTp \right)\right] \frac{Q'}{q} \right\} \Xi - 2 q
\varphi Q \varSigma - q \varphi \partial_\theta \varDelta  + q^2 \varphi^2
\GperT = 0,
\end{eqnarray}
where use has been made of Eq.~(\ref{eq:maxred}) to simplify the term
otherwise proportional to $\Omega$.

\subsection{Perturbed Klein-Gordon equation}

In the same way, the Klein-Gordon equation (\ref{eq:KG}) can also be
perturbed in terms of metric and matter fields.  By means of
Eqs.~(\ref{eq:perthiggs}) and (\ref{eq:complex_chi}), its real and
imaginary parts lead to two coupled equations,
\begin{eqnarray}
\label{eq:pertKG_real}
\ee^{-\sigma} \Box \varSigma & - & \varSigma'' - \left(2\sigma' +
\lngTTp \right) \varSigma' - \gTTi \partial_\theta^2 \varSigma + 2
\gTTi Q \partial_\theta \varDelta + \left[ \gTTi Q^2 +
\frac{\lambda}{2} \left(3\varphi^2 - \eta^2 \right) \right] \varSigma
\nonumber \\ & - & \varphi' \left(2\Psi' + \frac{\Omega'- \Xi'}{2} -
\gTTi \partial_\theta \Upsilon \right) + \left[ \varphi'' + \left(2
\sigma' + \lngTTp \right) \varphi' \right] \Xi - \gTTi
\varphi Q^2 \Omega - 2 \gTTi q \varphi Q \Theta_\theta =0,
\end{eqnarray}
and,
\begin{eqnarray}
\label{eq:pertKG_imag}
\ee^{-\sigma} \Box \varDelta & - & \varDelta'' - \left(2\sigma' +
\lngTTp \right) \varDelta' - \gTTi \partial_\theta^2 \varDelta - 2
\gTTi Q \partial_\theta \varSigma + \left[ \gTTi Q^2 +
\frac{\lambda}{2} \left(\varphi^2 - \eta^2 \right) \right] \varDelta
\nonumber \\ & + & \varphi Q \gTTi \left[\Upsilon' - \partial_\theta
\left( 2 \Psi + \dfrac{\Xi-\Omega}{2} \right) \right] + q \varphi
\left(-\ee^{-\sigma} \Box\Theta + \Theta_r' + \gTTi
\partial_\theta\Theta_\theta \right) \nonumber \\ & + &  \left[
\varphi Q' + 2 \varphi' Q + \varphi Q \left(2 \sigma' - \frac{1}{\rdim}
-\frac{\gamma'}{2} \right) \right] \gTTi \Upsilon + \left[2 \varphi' +
\left(2 \sigma' + \lngTTp \right) \varphi \right] q \Theta_r  =
0,
\end{eqnarray}
respectively.
\end{widetext}

As previously noted, the perturbed fields and geometry have to be
invariant by a complete rotation around the string, \ie they are
$2\pi$-periodic in the angular variable $\theta$. Therefore, they can
be decomposed in Fourier series with respect to $\theta$, \eg the
perturbed Higgs fields is expanded as
\begin{equation}
\label{eq:higgsfourier}
\begin{aligned}
\varDelta(r,\theta) & = \sum_{p \in \mathbb{Z}} \varDelta_p(r)
\ee^{ip\theta},\\ \varSigma(r,\theta) & = \sum_{p \in \mathbb{Z}}
\varSigma_p(r) \ee^{ip\theta},
\end{aligned}
\end{equation}
and similarly for all the other perturbations. Plugging
Eq.~(\ref{eq:higgsfourier}), and analogous mode expansion for $\Psi$,
$\Xi$, $\Omega$, $\Theta$, $\Theta_r$ and $\Theta_\theta$ into
Eqs.~(\ref{eq:einstein_munu}) to (\ref{eq:pertKG_imag}) clearly shows
that each angular ``$p$--mode'' decouples from the others. As a
result, the time evolution of the perturbations can be focused on a
particular angular mode $p$, the physical evolution ending up be given
by their superposition. Although some redefinitions of the fields may
separate the system of equations (\ref{eq:einstein_munu}) to
(\ref{eq:pertKG_imag}) into distinct
subsets~\cite{Randjbar-Daemi:2002pq}, such a situation already happens
for the lowest angular mode $p=0$. Indeed, the angular dependency of
these zero modes disappear which is formally equivalent to nullify the
``$\partial_\theta$'' operator in Eqs.~(\ref{eq:einstein_munu}) to
(\ref{eq:pertKG_imag}). One gets two disjoint pieces of equations,
namely Eqs.~(\ref{eq:einstein_munu}), (\ref{eq:einstein_mur}),
(\ref{eq:einstein_rr}), (\ref{eq:einstein_thetatheta}),
(\ref{eq:pertmax_theta}) and (\ref{eq:pertKG_real}) which only involve
$\Psi$, $\Omega$, $\Xi$, $\Theta_\theta$ and $\varSigma$ on one side,
while Eqs.~(\ref{eq:einstein_mutheta}), (\ref{eq:einstein_rtheta}),
(\ref{eq:pertmax_mu}), (\ref{eq:pertmax_r}) and (\ref{eq:pertKG_imag})
couple only $\Upsilon$, $\Theta_r$, $\Theta$ and $\varDelta$ on the
other side.

In fact the zero angular momentum modes, obtained for $p=0$, represent
cylindrical perturbations which strictly wind around the string as the
background forming fields do. Moreover, since they correspond to the
lowest angular momentum state, one may naturally expect them to be
first excited in a generic modification of the vortex structure.

In order to study the stability of these zero-modes, the time
evolution of the latter subset of perturbations will be thoroughly
analyzed in the following section.

\subsection{Stability of the transverse perturbations}
\label{sec:trans}
The time evolution of the lowest angular momentum modes
$\Upsilon_\zero$, $\Theta_\zero$, $\Theta_{r \zero}$ and
$\varDelta_\zero$ is readily governed by the perturbed Einstein
equations (\ref{eq:einstein_mutheta}) and (\ref{eq:einstein_rtheta})
together with the perturbed Maxwell equations (\ref{eq:pertmax_mu})
and (\ref{eq:pertmax_r}), and the perturbed Higgs one
(\ref{eq:pertKG_imag}). As noted before, due to implicit stress-energy
tensor conservation in the Einstein equations, this system is not
over-determined although it involves redundant equations. Moreover,
since we are interested in perturbations which behave almost like the
background vortex fields, we will only consider real perturbations of
the hyperstring forming Higgs field, \ie with $\varDelta_\zero=0$ (in
other words, we go to the real gauge). In terms of the dimensionless
background fields and parameters [see Eqs.~(\ref{eq:ansatz}),
(\ref{eq:dimless1}) and (\ref{eq:dimless2})], together with the new
dimensionless fields
\begin{equation}
\Thetadim = q \Theta, \quad \Thetadim_r = \frac{q}{\sqrt{|\Lambda|}}
\Theta_r, \quad \Upsiadim = \sqrt{|\Lambda|}\Upsilon ,
\end{equation}
the time evolution equations stemming from the Einstein and Maxwell
equations read
\begin{eqnarray}
\label{eq:closegauge_mutheta}
\! \! \! \dot{\Upsiadim} + \! \left(s + \frac{l}{2} \right)\Upsiadim -
\frac{4 \alpha \dot{Q}}{\varepsilon} \left( \Thetadim_r - \dot{\Thetadim}
\right) + 4 \alpha f^2 Q \Thetadim & = 0,& \\
\label{eq:closegauge_rtheta}
\left(\ee^{-\sigma} \madim^2 - 4 \alpha \frac{f^2 Q^2}{m}
\right) \Upsiadim - 4 \alpha f^2 Q \Thetadim_r & = 0,&\\
\label{eq:closegauge_mu}
\dot{\Thetadim}_r - \ddot{\Thetadim} + \left(s + \frac{l}{2} \right)
\left(\Thetadim_r - \dot{\Thetadim} \right) + \varepsilon f^2 \Thetadim
 &= 0,&\\
\label{eq:closegauge_r}
\ee^{-\sigma} \madim^2 \left(\Thetadim_r -\dot{\Thetadim} \right) -
\varepsilon \frac{f^2 Q}{m} \Upsiadim - \varepsilon f^2
\Thetadim_r & = 0,&
\end{eqnarray}
while the Higgs one becomes
\begin{eqnarray}
\label{eq:closehiggs_imag}
 \dfrac{Q}{m} \dot{\Upsiadim}  & + & \ee^{-\sigma} \madim^2 \Thetadim +
\dot{\Thetadim}_r +\left[ \dot{Q} + Q\left(2 \dfrac{\dot{f}}{f} +
2s-\dfrac{l}{2}\right) \right] \dfrac{\Upsiadim}{m} \nonumber \\
& +& \left(2 \dfrac{\dot{f}}{f} + 2s + \dfrac{l}{2} \right)\Thetadim_r  = 0.
\end{eqnarray}
A four-dimensional Fourier transform has been performed on the zero
angular momentum perturbed fields with respect to the four-dimensional
coordinates $x^\mu$. The general perturbation solution is therefore a
linear superposition of the d'Alembertian eigenmodes
\begin{equation}
\Upsiadim_\zero(x^\mu,r)=\int{\Upsiadim(k^\mu ,r) \ee^{-ik_\mu
x^\mu} \dd^4 k},
\end{equation}
with $\Upsiadim(k^\mu,r)$ the solution of
Eqs.~(\ref{eq:closegauge_mutheta}) to (\ref{eq:closehiggs_imag}), and
similarly for the others perturbed quantities. The d'Alembertian
eigenvalues ends up being
\begin{equation} \label{dAl}
\Box \longrightarrow -|\Lambda| \madim^2 = -\eta^{\mu
\nu}k_{\mu}k_{\nu},
\end{equation}
and any perturbation with positive mass squared $\madim^2\geq 0$ will
be considered stable, whereas tachyonic modes, having $\madim^2<0$,
will generate instabilities.

There are three variables $\Thetadim_r$, $\Thetadim$ and $\Upsiadim$
for five equations, two of them being thus constraint equations. By
means of Eq.~(\ref{eq:closegauge_rtheta}), the metric perturbation
$\Upsiadim$ can be expressed in terms of $\Thetadim_r$ only
\begin{equation}
\label{eq:upsitheta_r}
\Upsiadim = \dfrac{4 \alpha f^2 Q }{\ee^{-\sigma}
\madim^2 - 4 \alpha \dfrac{f^2 Q^2}{m}}\,\Thetadim_r ,
\end{equation}
while by means of Eq.~(\ref{eq:upsitheta_r}), Eq.~(\ref{eq:closegauge_r})
gives the relation
\begin{equation}
\label{eq:theta_r_theta}
\Thetadim_r  = (\calP + 1) \dot{\Thetadim}  ,
\end{equation}
with
\begin{equation}
\label{eq:pdef}
\calP = \dfrac{\varepsilon f^2}{\ee^{-\sigma} \madim^2 - \varepsilon
f^2 - 4 \alpha \dfrac{f^2 Q^2}{m}}.
\end{equation}
Finally, plugging the previous expressions for $\Upsiadim$ and
$\Thetadim_r$, given by Eq.~(\ref{eq:upsitheta_r}) and
Eq.~(\ref{eq:theta_r_theta}), into Eq.~(\ref{eq:closegauge_mu}), one
gets a second order differential equation involving only the
function $\Thetadim$, namely
\begin{equation}
\label{eq:evolthetadim}
\calP \ddot{\Thetadim} + \left[ \dot{\calP} +
\left(s + \dfrac{l}{2} \right) \calP \right]
\dot{\Thetadim} + \varepsilon f^2 \Thetadim = 0.
\end{equation}
One can also verify that the two remaining equations
(\ref{eq:closegauge_mutheta}) and (\ref{eq:closehiggs_imag}) also lead
to Eq.~(\ref{eq:evolthetadim}) when use is made of
Eqs.~(\ref{eq:upsitheta_r}) and (\ref{eq:theta_r_theta}), ensuring the
consistency of the gauge choice $\varDelta_\zero=0$. If this choice is
relaxed, one can check that the new perturbed equations simply require
$\Thetadim$ to be replaced by $\Thetadim - \varDelta_\zero/\varphi$,
and $\Thetadim_r$ by $\Thetadim_r -
\partial_\radim(\varDelta_\zero/\varphi)$ in
Eqs.~(\ref{eq:upsitheta_r}) to (\ref{eq:pdef}). The additional
perturbed equation stemming from Eq.~(\ref{eq:pertKG_imag}) ends up be
equivalent to Eq.~(\ref{eq:evolthetadim}). As expected for a gauge
degree of freedom, the field $\varDelta_\zero$ has therefore no
dynamics and will not be considered in the following.

In order to conclude on the stability of the vortex solution with
respect to these perturbations, let us consider the generic case of a
real squared mass $\madim^2\in\setR$. In this case, and far from the
string, Eq.~(\ref{eq:evolthetadim}) behaves as
\begin{equation}
\label{eq:evolthetaasymp}
\ddot{\Thetadim} - \sqrt{\dfrac{5}{2}} \dot{\Thetadim} + \madim^2
\exp{\left(\sqrt{\dfrac{2}{5}} \radim\right)} \Thetadim
\underset{\infty}{\sim} 0,
\end{equation}
and through the change of variable $z = \exp (\radim/\sqrt{10})$ and
function $\Thetadim = \exp(\sqrt{5/8}\radim) \tilde \calT$,
Eq.~(\ref{eq:evolthetaasymp}) reads
\begin{equation}
\frac{1}{z} \frac{\dd}{\dd z} \left( z \frac{\dd \tilde{\calT}}{\dd z}\right)
+ \left( {10 \madim^2} - \frac{25}{4 z^2} \right)
\tilde{\calT} =0,
\end{equation}
whose solutions are known, see, \eg Eq.~(8.491) in
Ref.~\cite{Gradshteyn:1965aa}. This gives
\begin{equation}
\label{eq:thetadiv} \Thetadim \propto \exp\left( \sqrt{\frac{5}{8}}
\radim \right) \times \calZ_{5/2} \left[ \sqrt{10} |\madim| \exp
\left( \frac{\radim}{\sqrt{10}}\right) \right] ,
\end{equation}
in which $\calZ_{5/2}$ is a Bessel function of order $5/2$ and of its
argument in brackets for $\madim^2>0$, and a modified Bessel function
for $\madim^2<0$. As a result, for any positive mass squared,
$\madim^2 >0$, the solution given in Eq.~(\ref{eq:thetadiv}) behaves,
asymptotically far from the vortex, as an oscillatory exponentially
divergent quantity whose amplitude scales as
\begin{equation}
|\Thetadim| \underset{\infty}{\propto} \exp{\left(\sqrt{\frac{2}{5}}
\radim \right)}.
\end{equation}
Such solutions appear to be unbounded far from the string and are,
strictly speaking, not well-defined: the first order perturbation
equations can be derived from the action expanded to second order in
these perturbations, which one would thus expect to be finite, since
the volume of the extra dimensions is finite (this is the very reason
for choosing anti-de Sitter in the first place). But this action
contains a term $\propto\int \sqrt{-g} \ee^{-\sigma} \Thetadim^2$ (the
exponential stemming from the unperturbed contravariant metric
coefficient) which diverges exponentially. In spite of this issue, the
solutions (\ref{eq:thetadiv}) can be given a physical meaning in the
framework of perturbation theory since the energy they contain to
first order is actually finite, and the corresponding gravitational
potential $\Upsiadim$ vanishes asymptotically [see
Eq.~(\ref{eq:upsitheta_r})]. In this respect, these positive squared
mass perturbations are physically admissible solutions.

In the case $\madim^2 < 0$, the Bessel function in
Eq.~(\ref{eq:thetadiv}) is of the modified kind and admits an
asymptotically exponential of exponential decaying behavior: one of
the two degrees of freedom of Eq.~(\ref{eq:thetadiv}) has to be fixed
to ensure the decrease of the $\madim^2 < 0$ solution
asymptotically. However, this does not mean that tachyonic modes exist
inside the system since it is also necessary that these perturbations
are well-defined in the string core. We shall accordingly turn
attention to the interior solution.

In the hyperstring core, the function $\calP$ can be expanded as
\begin{equation}
\calP = \dfrac{\varepsilon f_\one^2}{\madim^2 - 4 \alpha f_\one^2}
\radim^2 + \calO(\radim^4), 
\end{equation}
where use has been made of the background field behaviors in the
string core. In the limit $\radim \rightarrow 0$,
Eq.~(\ref{eq:evolthetadim}) becomes
\begin{equation}
\radim^2 \ddot{\Thetadim} + 3 \radim \dot{\Thetadim} + \radim^2
\left(\madim^2 - 4 \alpha f_\one^2 \right) \Thetadim
\simeq 0,
\end{equation}
whose solutions are
\begin{equation}
\Thetadim \propto \dfrac{1}{\radim} \calZ_1\left(\sqrt{|\madim^2 -
4\alpha f_\one^2|} \radim \right).
\end{equation}
Again, $\calZ_1$ refers to the two independent Bessel functions $J_1$
and $Y_1$ provided $\madim^2 - 4\alpha f_\one^2>0$, while it designs
the two modified one, $I_1$ and $K_1$, in the other case (which is the
case for $\madim^2<0$ in which we are interested). Therefore, there
always is a divergent solution in the string core, behaving as
$1/\radim^2$, together with a well-defined one going to a constant
value. As a result, the $\madim^2 < 0$ decreasing solution far from
the string can not generically match with the well-defined one in the
string core, the required degree of freedom being already fixed to
ensure the asymptotic normalizability. Although there is thus no
tachyonic continuum spectrum, the matching between the two
well-defined solutions at infinity and in the string core could happen
for some peculiar values of the negative mass squared, making a
discrete spectrum of unstable modes. In the following, we show that
this is not the case.

The equation (\ref{eq:evolthetadim}) can be rewritten in the form of a
zero mode Schr\"odinger equation for the quantity $u\equiv
\left[\radim|\calP|\exp\left(\sigma+\gamma/2\right)
\right]^{1/2}\Thetadim$, namely
\begin{equation}
\label{eq:Schrod}
-\ddot u + V_M (\radim) u=0,
\end{equation}
with the potential
\begin{equation}
\label{eq:Schrodu}
V_M (\radim)= W^2 +\dot W +\varepsilon f^2+4\alpha\frac{f^2Q^2}{m} -
\madim^2 \ee^{-\sigma},
\end{equation}
with 
\begin{equation}
W=\frac{1}{2}\left[\frac{\dot\calP}{\calP}+
\left(s+\frac{l}{2}\right)\right],
\label{eq:W}\end{equation}
the superpotential-like of the Schr\"odinger
equation~(\ref{eq:Schrod})~\cite{Cooper:1994eh}.

\begin{figure}
\begin{center}
\includegraphics[width=8.5cm,angle=0]{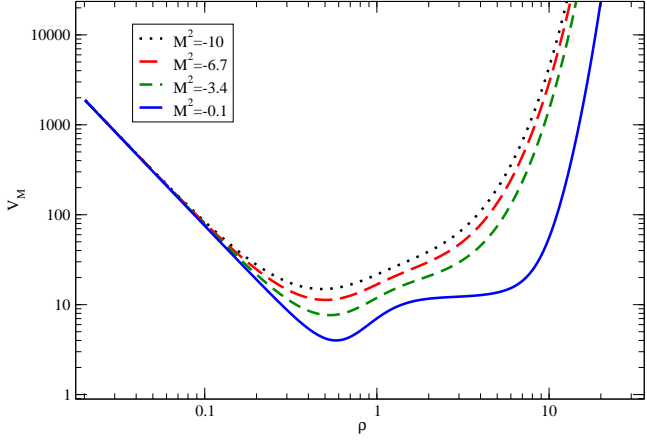}
\caption{Characteristic shape of the potential $V_M(\radim)$ given by
Eq.~(\ref{eq:Schrodu}) for a typical set of values
$(\alpha,\beta,\varepsilon)=(1.77, 2.66, 11.56)$ on the
fine-tuning surface for which $v_\zero =1$ and different values of the
negative squared mass $\madim^2$. It is clear from this figure that
the potential is positive definite, and its minimum increases with
$|\madim^2|$. Hence, there is no zero mode for the perturbation
equation, and therefore no instability zone.}
\label{fig:Vu}
\end{center}
\end{figure}

It is immediately clear from Eq.~(\ref{eq:Schrodu}) that the potential
is asymptotically dominated by the last, exponentially increasing,
term. As a result, a confinement for the corresponding scalar mode is
achieved provided this last term is positive, hence requiring a
negative squared mass. This would seem to imply the existence of
tachyonic modes on the hyperstring. However, a closer examination of
the potential (displayed on Fig.~\ref{fig:Vu}) actually shows that
even in the negative squared mass case, $V_M$ is, numerically,
positive definite: the associated Schr\"odinger equation has only
strictly positive eigenvalues, and in particular no zero mode. This
can be seen analytically in the following way: we first note that
Eq.~(\ref{eq:Schrod}) can be written as $\left[ \mathcal{A}^\dagger
\mathcal{A} + Z^2(\radim) \right] u =0$, where $Z$ can be obtained
from the potential (\ref{eq:Schrodu}) and the definition of the
operator
\begin{equation}
\mathcal{A} \equiv \frac{\dd}{\dd \radim} + W (\radim), \quad \to
\quad \mathcal{A}^\dagger \equiv -\frac{\dd}{\dd \radim} + W
(\radim),
\label{eq:AA+}
\end{equation}
\ie
$$Z^2 = \varepsilon f^2 + \frac{4\alpha f^2Q^2}{m}-\madim^2
\ee^{-\sigma},$$ which is indeed a positive definite function of the
core distance. Then, if one demands the perturbation $u$ to be
bounded, as implied by the fact that the energy contained in the mode
be finite, then one is seeking zero mode bound (normalizable) states
of the Schr\"odinger equation (\ref{eq:Schrod}). It is however
immediately clear that since the operator
$\mathcal{A}^\dagger\mathcal{A}$ has only nonnegative eigenvalues,
then the spectrum of $\mathcal{A}^\dagger \mathcal{A} + Z^2(\radim)$
must be positive definite; thus, as announced, there are no zero modes
solution of Eq.~(\ref{eq:Schrod}), and hence no instability. It is
interesting to notice that the nonnormalizability property at the
hyperstring core also stems from the remark that the potential
(\ref{eq:Schrodu}) diverges as $V_M\sim\radim^{-2}$ near the core.

Thus, the tachyonic solutions cannot be considered as physical
perturbations of the system. We are led to the conclusion that, for
the subset of fields considered in this section, the hyperstring is
stable with respect to transverse perturbations. To decide on the
overall stability of the vortex, one needs to clarify the role of the
complementary subset of modes; This is done in the following section.

\subsection{Sausage perturbation modes}
\label{sec:saus}
We now turn to the second decoupled subset of cylindrical perturbation
modes which respect the axial symmetry, hence their name. Provided we
are interested in the zero angular momentum modes, the time evolution
of the perturbations $\Psi$, $\Xi$, $\Omega$, $\Sigma$ and
$\Theta_\theta$ is determined by the Eqs.~(\ref{eq:einstein_munu}),
(\ref{eq:einstein_mur}), (\ref{eq:einstein_rr}),
(\ref{eq:einstein_thetatheta}), (\ref{eq:pertmax_theta}) and
(\ref{eq:pertKG_real}). Some rapid simplifications can be
performed. First, from the $\mu \ne \nu$ part of the Einstein equation
(\ref{eq:einstein_munu}), one gets
\begin{equation}
\label{eq:omega}
\Omega = -2 \Psi - \Xi.
\end{equation}
This expression can thus be used to simplify the above mentioned
equations for $\mu=\nu$. In terms of dimensionless quantities the
perturbed Einstein equations simplify into two dynamical equations
\begin{align}
\label{eq:close_mumu}
\ddot{\Psi}  &- \ddot{\Xi} + \left(3s -\dfrac{l}{2} \right) \dot{\Psi} -
\left(3s + \dfrac{3}{2}l \right) \dot{\Xi} \nonumber \\
& + \left(2\calV + 2\calVV
 \right) \Psi + \left(2 \calF + \calV + 2 \calVV \right) \Xi =
\calS_{\mu \mu},\\
\label{eq:close_thetatheta}
4 \ddot{\Psi} &+ 10s \dot{\Psi} - 2s \dot{\Xi} + \left(3
\ee^{-\sigma} \madim^2 -2 \calV -2 \calVV \right)\Psi \nonumber \\
&+ \left(\ee^{-\sigma} \madim^2 + 2 \calF -\calV -2\calVV \right) \Xi
= \calS_{\theta \theta},
\end{align}
and two constraint equations
\begin{align}
\label{eq:close_mur}
\dot{\Psi}  - \dot{\Xi} &+ (s - l) \Psi - (s + l) \Xi  = \calS_{\mu r},
\\
\label{eq:close_rr}
2(s+l) \dot{\Psi}  &-  2s \dot{\Xi} + \left(\ee^{-\sigma} \madim^2 -
2 \calV + 2 \calVV \right) \Psi \nonumber \\
&+ \left(-\ee^{-\sigma}\madim^2 + 2\calF -\calV + 2\calVV \right) \Xi
=\calS_{rr},
\end{align}
where the matter source field functions $\calS$ stand for
\begin{align}
\label{eq:src_mumu}
\calS_{\mu \mu} & = 4 \alpha \dfrac{Q\dot{Q}}{\varepsilon m}
\dot{\Thetadim}_\theta - 4 \alpha f \dot{f} \dot{\Sigmadim} + 2 (\calV
+ \calVV) \Thetadim_\theta \nonumber \\ &- \left[4 \alpha \dot{f}^2 +
2\calVV + 2f \dfrac{\dd \calF}{\dd f} \right]\Sigmadim,
\end{align}
\begin{align}
\label{eq:src_thetatheta}
\calS_{\theta \theta} &= -4 \alpha \dfrac{Q\dot{Q}}{\varepsilon
m}\dot{\Thetadim}_\theta - 4 \alpha f \dot{f} \dot{\Sigmadim} -
2(\calV + \calVV) \Thetadim_\theta \nonumber \\ &- \left[4\alpha \dot{f}^2 -
2 \calVV + 2f \dfrac{\dd \calF}{\dd f} \right]\Sigmadim,
\end{align}
\begin{align}
\label{eq:src_mur}
\calS_{\mu r} &= \dfrac{4\alpha Q \dot{Q}}{ \varepsilon m}
\Thetadim_\theta - 4 \alpha f \dot{f} \Sigmadim,
\end{align}
\begin{align}
\label{eq:src_rr}
\calS_{rr} &= -4\alpha \dfrac{Q\dot{Q}}{\varepsilon m}
\dot{\Thetadim}_\theta + 4 \alpha f \dot{f} \dot{\Sigmadim} -2(\calV -
\calVV) \Thetadim_\theta \nonumber \\ & - \left[-4 \alpha \dot{f}^2 +
2 \calVV + 2f \dfrac{\dd \calF}{\dd f} \right]\Sigmadim,
\end{align}
in which $\Thetadim_\theta \equiv q\Theta_\theta/Q$ and $\Sigmadim
\equiv \Sigma/\varphi$.

The perturbed Klein-Gordon and Maxwell equations, also expressed in
terms of dimensionless quantities read
\begin{align}
\label{eq:closehiggs_real}
\ddot{\Sigmadim} & + \left(2\dfrac{\dot{f}}{f}+ 2s + \dfrac{l}{2}
\right) \dot{\Sigmadim} + \left( \ee^{-\sigma}\madim^2 - 4 \beta f^2
\right) \Sigmadim \nonumber \\
&= \dfrac{\dot{f}}{f}\left(\dot{\Xi} -
\dot{\Psi}\right) + 2 \left[\dfrac{Q^2}{m} + \beta f(f^2-1) \right]
\Xi  \nonumber \\
&+ 2 \dfrac{Q^2}{m} \left(\Psi -\Thetadim_\theta\right),
\end{align}
and
\begin{align}
\label{eq:closegauge_theta}
\ddot{\Thetadim}_\theta &+ \left(2\dfrac{\dot{Q}}{Q} + 2s -
\dfrac{l}{2} \right) \dot{\Thetadim}_\theta + \ee^{-\sigma} \madim^2
\Thetadim_\theta \nonumber \\ & = 3\dfrac{\dot{Q}}{Q} \dot{\Psi}
-\varepsilon f^2 \left(\Xi +2 \Sigmadim\right).
\end{align}
It is interesting to note at this point that the equations of motion
for the five-dimensional domain wall having this kind of perturbation
modes are directly obtained from Eqs.~(\ref{eq:close_mumu}) to
(\ref{eq:closehiggs_real}) by setting $\Omega=0$ and removing any
dependencies in $l$ and $Q$.

As previously noted, due to the Bianchi identities some of these
equations are redundant. Indeed, Eq.~(\ref{eq:close_mumu}) is readily
obtained by differentiation of Eq.~(\ref{eq:close_mur}), up to the
background Einstein, Klein-Gordon and Maxwell equations (\ref{eq:s})
to (\ref{eq:w}). Similarly, differentiating Eq.~(\ref{eq:close_rr}),
and using Eqs.~(\ref{eq:close_mur}), (\ref{eq:closehiggs_real}) and
(\ref{eq:closegauge_theta}) to express $\dot{\calS}_{rr}$ in terms of
$\calS_{\theta\theta}$ and the metric perturbations leads to
Eq.~(\ref{eq:close_thetatheta}), also up to the background Einstein,
Klein-Gordon and Maxwell equations, and provided $l \ne 0$. As a
result, only Eqs.~(\ref{eq:omega}), (\ref{eq:close_mur}),
(\ref{eq:close_rr}), (\ref{eq:closehiggs_real}) and
(\ref{eq:closegauge_theta}) are relevant for this subset of matter and
metric perturbations, with the constraint that the solutions are
regular at the point where $l$ vanishes. However this system remains
fully coupled and no simple second order differential equation on one
perturbation variable can be obtained, as it was the case for the
transverse modes [see Eq.~(\ref{eq:evolthetadim})].

To study the stability properties of the hyperstring with respect to
the sausage modes we derive in the following their behavior in the
string core and asymptotically.

\begin{widetext}
The previous system can be recast into a set of first order
differential equations,
\begin{align}
\label{eq:psidot}
2 l \dot{\Psi} &= \left[- \ee^{-\sigma} \madim^2 + 2 \calV - 2 \calVV
+ 2s(s-l) \right] \Psi + \left[\ee^{-\sigma} \madim^2 + \calV - 2
\calVV -2 \calF -2s(s+l) \right] \Xi \nonumber \\ &+ \left[4 \alpha f
\dot{f}\left(\dfrac{\dot{f}}{f} + 2s\right) - 2\calVV -2f \dfrac{\dd
\calF}{\dd f}\right] \Sigmadim + 4 \alpha f \dot{f} \Sigmadimdot +
\big[-2\calV + 2 \calVV -2s(s-l) \big] \Thetadim - (s-l)
\Thetadimdot,\\
\label{eq:xidot}
2l \dot{\Xi} &= \left[- \ee^{-\sigma} \madim^2 + 2 \calV - 2 \calVV +
2(s+l)(s-l) \right] \Psi + \left[\ee^{-\sigma} \madim^2 + \calV - 2
\calVV -2 \calF -2(s+l)^2 \right] \Xi \nonumber \\ &+ \left[4 \alpha f
\dot{f} \left(\dfrac{\dot{f}}{f} + 2s +2l \right) -2 \calVV -2f
\dfrac{\dd \calF}{\dd f} \right] \Sigmadim + 4 \alpha f \dot{f}
\Sigmadimdot + \big[-2 \calV + 2 \calVV -2(s+l)(s-l) \big]
\Thetadim - (s-l) \Thetadimdot,
\end{align}
for the metric perturbations
\begin{align}
\label{eq:sigmadot}
\dot{\Sigmadim} &= \Sigmadimdot,\\
\label{eq:sigmaddot}
\dot{\Sigmadimdot} &= \left[2 \dfrac{Q^2}{m} + \dfrac{\dot{f}}{f}(s-l)
\right] \Psi + \left[2 \dfrac{Q^2}{m} - \dfrac{\dot{f}}{f} (s+l) +
\dfrac{1}{2 \alpha} \dfrac{\dd \calF}{\dd f} \right] \Xi +
\left[-\ee^{-\sigma} \madim^2 + 4 \alpha \dot{f}^2 + 4 \beta f^2
\right] \Sigmadim \nonumber \\
&- \left[2 \dfrac{\dot{f}}{f} + 2s + \dfrac{1}{2}l
\right] \Sigmadimdot - \left[2 \dfrac{Q^2}{m} +
\dfrac{\dot{f}}{f}(s-l)\right] \Thetadim,
\end{align}
for the Higgs field perturbations and
\begin{align}
\label{eq:thetadot}
\dot{\Thetadim} &= \Thetadimdot,\\
\label{eq:thetaddot}
2l \dot{\Thetadimdot} &= 3 \dfrac{\dot{Q}}{Q} \left[- \ee^{-\sigma}
\madim^2 + 2 \calV -2 \calVV +2s(s-l) \right] \Psi + \left\{-
2\varepsilon l f^2 + 3 \dfrac{\dot{Q}}{Q} \left[\ee^{-\sigma} \madim^2
+ \calV -2 \calVV -2 \calF -2s(s+l) \right] \right\} \Xi \nonumber \\
&+ \left\{ -4 \varepsilon l f^2 + 3 \dfrac{\dot{Q}}{Q} \left[4 \alpha
f \dot{f} \left(\dfrac{\dot{f}}{f} + 2s \right) - 2 \calVV -2f
\dfrac{\dd \calF}{\dd f} \right] \right \} \Sigmadim + 12 \alpha
\dfrac{\dot{Q}}{Q} f \dot{f} \Sigmadimdot \nonumber \\ &+ \left\{ -2l
\ee^{-\sigma} \madim^2 + 3\dfrac{\dot{Q}}{Q} \left[-2 \calV + 2 \calVV
-2s(s-l)\right] \right\} \Thetadim - \left[\dfrac{\dot{Q}}{Q}(3s+l) +
2l(2s - \dfrac{1}{2} l) \right] \Thetadimdot,
\end{align}
for the gauge field.
\end{widetext}

{}From Eqs.~(\ref{eq:psidot}), (\ref{eq:xidot}) and making use of the
asymptotic behaviors of the background fields (see
Sect.~\ref{Sec:IV}), the metric sausage perturbation modes at infinity
verify
\begin{align}
\label{eq:psidotinfty}
\dot{\Psi} & \simeq \sqrt{\dfrac{5}{2}} \ee^{\sqrt{2/5} \radim}
\madim^2 \dfrac{\Psi}{2} - \left(\sqrt{\dfrac{5}{2}} \ee^{\sqrt{2/5}
\radim} \madim^2 + \sqrt{\dfrac{2}{5}} \right) \dfrac{\Xi}{2},
\end{align}
and
\begin{align}
\label{eq:xidotinfty}
\dot{\Xi} & \simeq \sqrt{\dfrac{5}{2}} \ee^{\sqrt{2/5} \radim}
\madim^2 \dfrac{\Psi}{2} - \left(\sqrt{\dfrac{5}{2}} \ee^{\sqrt{2/5}
\radim} \madim^2 - 3\sqrt{\dfrac{2}{5}} \right) \dfrac{\Xi}{2}.
\end{align}
{}From Eq.~(\ref{eq:psidotinfty}), one gets
\begin{equation}
\label{eq:xiinfty}
\Xi \simeq \dfrac{\ee^{\sqrt{2/5} \radim} \madim^2 \Psi - 2 \sqrt{2/5}
\dot{\Psi}}{\ee^{\sqrt{2/5} \radim} \madim^2 + 2/5},
\end{equation}
while Eq.~(\ref{eq:xidotinfty}) yields
\begin{equation}
\label{eq:psiddotinfty}
\ddot{\Psi} - \sqrt{\dfrac{5}{2}} \dot{\Psi} + \ee^{\sqrt{2/5} \radim}
\madim^2 \Psi \simeq 0.
\end{equation}
{}From Eq.~(\ref{eq:evolthetaasymp}) and (\ref{eq:thetadiv}), the
metric perturbations $\Psi$, and therefore $\Xi$, diverge at infinity
as $\exp{(\sqrt{2/5} \radim)}$ as long as $\madim^2>0$. This behavior
is not admissible in the framework of perturbation theory. As can be
seen in Eq.~(\ref{eq:pertmetric}), the metric perturbations $\Psi$ and
$\Xi$ have to be small compared with their corresponding background
values, themselves are of order unity, at least initially. Otherwise,
mathematically speaking, it is not consistent to expand the equations
of motion to first order. Physically, this means we would start from a
space which is infinitely far from the background one. 

Similar conclusions also hold for the matter fields: {}from
Eqs.~(\ref{eq:sigmadot}) to (\ref{eq:thetaddot}), one gets
asymptotically for the Higgs field perturbations
\begin{equation}
\label{eq:sigmadimddotinfty}
\ddot{\Sigmadim} - \sqrt{\dfrac{5}{2}} \dot{\Sigmadim} +
\ee^{\sqrt{2/5} \radim} \madim^2 \Sigmadim \simeq 0,
\end{equation}
and, using Eq.~(\ref{eq:gaugeF}),
\begin{equation}
\label{eq:thetadimddotinfty}
\ddot{\Thetadim} - \left(2 \ellg + \dfrac{3}{2}
\sqrt{\dfrac{2}{5}}\right) \Thetadim + \ee^{\sqrt{2/5} \radim}
\madim^2 \Thetadim \simeq -3 \ellg \dot{\Psi},
\end{equation}
for the gauge field. For $\madim^2>0$, both of these equations have
only divergent behavior at infinity, as $\exp{(\sqrt{2/5} \radim)}$
for the Higgs perturbations and as $\exp{[(2 \ellg +
\sqrt{2/5})\radim/2]}$ for the gauge field perturbations [see
Eq.~(\ref{eq:epsilonmin})]. Also for the matter fields, the
$\madim^2>0$ solutions are not admissible since it would physically
means that the Higgs field is infinitely far from its vacuum
expectation value $f=1$.

As a result, the asymptotic study of the sausage modes shows that the
hyperstring cannot be stable with respect to these perturbations. As
this stage, either there are instabilities if there exist some
$\madim^2<0$ modes which are well-defined in the hyperstring core and
match the decreasing solution at infinity (see Sect.~\ref{sec:trans}),
or the configuration is not perturbable at all, \ie the only
acceptable solution is vanishing perturbations. To explore this point,
we discuss the behavior of the solutions in the hyperstring core.

The physical solutions we are interested in have to be well-defined at
$\radim=0$. In particular, the geometry can only be regular provided
$\dot{\Psi}(0) = \dot{\Xi}(0)=0$ [see Eq.~(\ref{eq:ricci}) and
discussion below for the background case]. Moreover, the sausage
perturbations are required to be small with respect to their
corresponding background values, \ie $\Psi$, $\Xi$, $\Sigmadim$ and
$\Thetadim$ have to be finite in the core. Assuming these fields can
be expanded in Laurent series around $\radim=0$, the previous
constraints yield
\begin{equation}
\label{eq:laurent}
\begin{aligned}
\Psi &\underset{0}{\sim} \psi_\zero + \sum_{n=2}^{\infty} \psi_n \radim^n, &
\Xi &\underset{0}{\sim} \xi_\zero + \sum_{n=2}^{\infty} \xi_n \radim^n, \\
\Sigmadim &\underset{0}{\sim} \sum_{n=0}^{\infty} \sigma_n \radim^n, &
\Thetadim &\underset{0}{\sim} \sum_{n=0}^{\infty} \theta_n \radim^n,
\end{aligned}
\end{equation}
where $\psi_n$, $\xi_n$, $\sigma_n$ and $\theta_n$ are real
numbers. Plugging these expansions into Eqs.~(\ref{eq:psidot}),
(\ref{eq:xidot}), (\ref{eq:sigmadot}), (\ref{eq:sigmaddot}),
(\ref{eq:thetadot}) and (\ref{eq:thetaddot}), where the derivatives
with respect to $\radim$ are readily obtained from
Eq.~(\ref{eq:laurent}), leads to a set of coupled algebraic relations
for the coefficients $\psi_n$, $\xi_n$, $\sigma_n$ and $\theta_n$. By
using the behaviors of the background fields in the hyperstring core
obtained in Sect.~\ref{sect:near}, we find that this hierarchy is fully
determined by the knowledge of the three parameters $\psi_0$,
$\sigma_0$ and $\theta_2$. As a result, the regular solutions in the
hyperstring core generate a three-dimensional subspace of the
six-dimensional full space of solutions. It is therefore necessary to
fix three degrees of freedom to get regular solutions in $\radim=0$.

{}From Eqs.~(\ref{eq:xiinfty}), (\ref{eq:psiddotinfty}),
(\ref{eq:sigmadimddotinfty}) and (\ref{eq:thetadimddotinfty}), we see
that three degrees of freedom must also be fixed to ensure that the
sausage perturbations are asymptotically well-defined (one for the
metric perturbations $\Psi$ and $\Xi$, one for the Higgs field
perturbation $\Sigmadim$ and one for the gauge field perturbation
$\Thetadim$). Moreover, from Eqs.~(\ref{eq:psidot}), (\ref{eq:xidot})
and (\ref{eq:thetaddot}), there will be no jump in the derivative of
the perturbations at $l=0$ provided
\begin{widetext}
\begin{equation}
\begin{aligned}
&\left(-\ee^{\sigma} \madim^2 + 2\calV -2\calVV +2s^2 \right)
\Psi(\radim_\uc) + \left(-\ee^{\sigma} \madim^2 + \calV -2\calVV -2\calF
-2s^2 \right) \Xi(\radim_\uc)  \\ & + \left[4 \alpha f
\dot{f}\left(\dfrac{\dot{f}}{f} + 2s\right) - 2\calVV -2f
\dfrac{\dd\calF}{\dd f}\right] \Sigmadim(\radim_\uc) + 4 \alpha f
\dot{f} \Sigmadimdot(\radim_\uc) + \left[-2\calV + 2 \calVV -2s^2 \right]
\Thetadim(\radim_\uc) - s \Thetadimdot(\radim_\uc) = 0,
\end{aligned}
\end{equation}
\end{widetext}
where the background fields are evaluated at $\radim=\radim_\uc$, the
vanishing point of $l$.

Three degrees of freedom are thus fixed to ensure a convergent
behavior of the perturbations at infinity, plus another one for the
regularity at $l=0$. Two degrees of freedom are left, which is not
sufficient, according to the previous discussion, to ensure regularity
of the solutions in the hyperstring core. Even if one tunes the mass
$\madim^2$ to keep only the solutions regular at $\radim=0$, the
convergent solutions at infinity (and regular at $l=0$) will not
generically match with the regular ones in the vortex core. We have
also numerically verified that no exceptional hidden symmetry realizes
this matching for a wide range of masses. Nevertheless, note that
another parameters of the model could be used to realize the matching
between the regular solutions in the core and at infinity. Indeed, the
background fine-tuning between $\alpha$, $\beta$ and $\varepsilon$
(see Fig.~\ref{fig:regsurf}) is a surface in the three-dimensional
parameters space and one cannot exclude that instabilities could
marginally occur on a curve along this surface.

In conclusion, the only acceptable sausage perturbations modes are the
vanishing ones, \ie the hyperstring cannot be perturbed at all in the
subset of matter and metric perturbations which correspond to
nonvanishing background fields. In the following, we discuss the
physical meaning of this result.

{}From a geometric point of view, it is well known that the generic
space-time generated by a cosmic string in presence of a cosmological
constant is of infinite-volume~\cite{Linet:1984} (see
Appendix~\ref{App:6dLambda} for a six-dimensional analogous
derivation). This is precisely why we have to fine tune the model
parameters to obtain a finite-volume space-time with decreasing warp
factors and no singularity in the core. As can be seen from the metric
(\ref{eq:metric}), the obtained space-time geometry leads to vanishing
proper length circles around the hyperstring at infinity. This kind of
geometry implies the existence of a point where $l(\radim_\uc)=0$
which is precisely the stationary point of proper length
circles. {}From $\radim < \radim_\uc$ the proper perimeter of a circle
around the hyperstring increases with respect to the radius, whereas
for $\radim>\radim_\uc$ it decreases toward $0$ (see
Fig.~\ref{fig:solgrav}). To link this structure to the usual conical
geometry generated by cosmic strings, one may imagine a missing angle
starting from zero in the hyperstring core toward $2 \pi$ at
infinity. This is not really surprising since we have required the
hyperstring to generate an anti-de Sitter space-time at infinity, or
naively, the fine-tuning allows to pass from a cylindrical symmetry in
the core to a spherical one asymptotically. Now, it is clear that
disturbing the fields around the values which lead to such fine-tuned
gravitational configuration is not necessary allowed. And this is
precisely our result. The only allowed perturbations concern the
transverse modes which have not equivalent at zero order. On the
contrary, all perturbations of the background fields, those generating
this fine-tuned space-time, are forbidden. Interestingly, the natural
behavior of the perturbations far the string for $\madim^2>0$, as
$\exp(\sqrt{2/5} \radim)$, looks like the generic metric coefficients
which appears when there is no fine-tuning (see
Appendix~\ref{App:6dLambda}). Although one might expect the
hyperstring to relax toward this generic configuration, no conclusion
can be drawn from the perturbations theory since the generic space-time
configuration, with infinite-volume, is not at all ``close to'' the
studied fine-tuned one. To end this section, it is worth pointing out
that the previous conclusion is valid in the physical motivated
framework where the vortex remains regular in $\radim=0$. This
hypothesis has allowed us to set the regular expansions in
Eq.~(\ref{eq:laurent}). On the other hand, as shown for the background
fields (see Sect~\ref{Sec:IV} and Sect.~\ref{Sec:V}), there is a dense
set of solutions associated with an anti-de Sitter space-time at
infinity which exhibit conical, or curvature singularities, in the
vortex core (see Fig.~\ref{fig:sing}). This suggest that an allowed
evolution of the fine-tuned regular vortex might be the birth of a
singularity in $\radim=0$ (see Appendix~\ref{App:6dLambda}).

\section{Conclusions}\label{Sec:VII}

In the braneworld framework, one issue is to determine how to model
the brane, and in particular to investigate whether its internal
structure influences the properties of gravity and of the other fields
living on the brane. Among other solutions, more interests has been
focused on the possibility for the brane to be realized by a
topological defect~\cite{Rubakov:1983bb,Visser:1985qm, Antunes:2002hn}.

In five dimensions, it has been found that there always exists a
domain wall solution that confines gravity, which moreover is
symmetric with respect to both sides of the brane provided the usual
relationship between the bulk and the brane cosmological constants is
satisfied. This relationship translates into a fine-tuning of the
underlying microphysics parameters~\cite{Ringeval:2001cq}. As far as
the gravitational sector is concerned, the properties of the
braneworld are mostly independent of the internal structure of the
brane. It is then possible to find various confinement mechanisms that
lead to the existence of bosonic~\cite{Bajc:1999mh, Dvali:1996bg,
Dvali:1997xe, Dubovsky:2000am, Dvali:2000rx, Dimopoulos:2000ej,
Duff:2000jk, Oda:2001ux, Ghoroku:2001zu, Akhmedov:2001ny} as well as
fermionic~\cite{Neronov:2001qv} zero modes that can be made
massive~\cite{Ringeval:2001cq, Dvali:2001qr} (although with a spectrum
not yet compatible with accelerator data). In short, a five
dimensional topological model of our Universe is, for the time being,
an open possibility both from the cosmological and particles physics
points of view.

In six dimensions, the general machinery used to study
five-dimensional reflection symmetric braneworld does not apply,
mainly because of the necessity to regularize the long range
gravitational self-interaction~\cite{Carter:2002tk}. A way around is
to specify a complete model determining the internal structure of the
brane in order to grasp some features of six-dimensional braneworld
models. Indeed, one will then need to discuss the genericness of the
conclusions drawn on a particular microphysics. For instance, there
exists a vortexlike brane configuration on which gravity was shown to
be localizable~\cite{Giovannini:2001hh,Giovannini:2002mk}. We have
shown in this article that such vortexlike branes are generically
associated with a singularity in the core, except when a fine-tuning
between the model parameters is assumed~\cite{Giovannini:2001hh}.

In order to address the stability issue of the fine-tuned solution, we
have performed a full gauge-invariant perturbation theory around the
regular six-dimensional vortex background solution. Focusing on the
scalar perturbations, we showed that the hyperstring forming fields
and nonvanishing metric parts cannot be perturbed at all. This result
comes from the requirement that the hyperstring generates a finite
volume space-time with infinite extra dimension and remains regular in
the core. Any perturbation of the background fields would destroy this
configuration and are not allowed. As a result, a nonempty universe,
where additional observable fields would generate such perturbations,
is severely constrained, if not altogether ruled out. Indeed all
induced modifications of the string and gauge forming fields, as time
and radial metric factors are forbidden, even at the perturbation
level. In other words, such nonempty universe should not couple to
Higgs and gauge fields, and should not modify at all the gravity in
the radial extra dimension. Therefore, the physical status of such a
configuration seems rather unclear, rather artificial to say the
least, and the possibility of having such a 6D vortex-brane
realization in nature very dubious.

A priori, these conclusions are specific to the case at hand. In
particular they might be argued to depend on the field content of the
underlying theory. However, since the discussion involved the
gravitational sector, it could be conjectured that six-dimensional
braneworld with anti-de Sitter bulk and infinite extra dimension,
cannot similarly be perturbed without exhibiting singularities. This
would imply that nonempty multi-dimensional braneworld models could
only have one (large) extra dimension of the warped form (unless some
extra structure, such as a 4-brane at a fixed and finite location away
from the 3-brane is added~\cite{Cline:2003ak}).

\acknowledgments We wish to thank Brandon Carter, Christos Charmousis,
Ruth Durrer, Massimo Giovannini, Seif Randjbar-Daemi, Mikhail
Shaposnikov and J\'er\'emie Vinet for enlightening discussions. We are
specially indebted to Harvey Meyer who explained in details the
numerical method used in Ref.~\cite{Giovannini:2001hh}.

\appendix

\section{Perturbed quantities}\label{App:dvp}

In this appendix, we derive all the gauge-invariant parts of the
various tensors necessary for the stability analysis.

\subsection{Metric tensor}
According to Eq.~(\ref{eq:pertmetric}), the scalar perturbed metric
tensor reads, in terms of gauge-invariant variables,
\begin{equation}
\begin{aligned}
\delta g_{\mu \nu} & = \ee^\sigma \eta_{\mu \nu} \Psi, & \delta g_{\mu
r} & = 0, & \delta g_{\mu \theta} & = 0, \\ \delta g_{r \theta} & =
-\Upsilon, & \delta g_{rr} & = - \Xi, & \delta g_{\theta \theta} & =
-\gTT \Omega.
\end{aligned}
\end{equation}
By means of
\begin{equation}
\delta g^{\si{AB}}=-g^{\si{AC}} g^{\si{BD}} \delta g_{\si{BD}},
\end{equation}
one can get the inverse perturbed metric tensor
\begin{equation}
\begin{aligned}
\delta g^{\mu \nu} & = - \ee^{-\sigma} \eta^{\mu \nu} \Psi, & \delta
g^{\mu r} & = 0, & \delta g^{\mu \theta} & = 0,\\ \delta g^{r \theta}
& = \gTTi \Upsilon, & \delta g^{rr} & = \Xi, & \delta g^{\theta
\theta} & = \gTTi \Omega.
\end{aligned}
\end{equation}
The perturbed Riemann tensor can also be expressed as a function of
the perturbed metric tensor through the perturbed Christoffel symbols
\begin{equation}
\delta R^{\si A}_{\si{\ BCD}} = -\delta \Gamma^{\si A}_{\si{\ BC};{\si
D}} + \delta \Gamma^{\si A}_{\si{\ BD};{\si C}},
\end{equation}
where the covariant derivatives with respect to the unperturbed metric
have been noted with a semicolon, and the perturbed connections are
given by
\begin{equation}
\delta \Gamma^{\si A}_{\si{\ BC}} = \frac{1}{2} g^{\si{AD}} \left(
\delta g_{\si{DB;C}} + \delta g_{\si{DC};{\si B}} - \delta
g_{\si{BC};{\si D}} \right).
\end{equation}

\subsection{Einstein tensor}

{}From the perturbed Riemann tensor, the perturbed Einstein tensor can
be expressed in terms of gauge-invariant variables by means of
\begin{equation}
\delta G_{\si{AB}} = \delta R_{\si{AB}} - \frac{1}{2} R \delta g_{\si{AB}} -
\frac{1}{2} g_{\si{AB}} \delta R,
\end{equation}
where the perturbed Ricci scalar is
\begin{equation}
\delta R = g^{\si{AB}} \delta R_{\si{AB}} + \delta g^{\si{AB}}
R_{\si{BD}}.
\end{equation}
After some (tedious) calculations one gets
\begin{widetext}
\begin{equation}
\begin{aligned}
\delta G_{\mu \nu} & = \left(\partial_\mu \partial_\nu - \eta_{\mu
\nu} \Box \right) \left( \frac{\Xi + \Omega}{2} + \Psi \right) +
\frac{1}{2} \ee^\sigma \eta_{\mu \nu} \Bigg\{3 \Psi'' + 3 \gTTi
\partial^2_\theta \Psi + \Omega'' - 2 \gTTi \partial_\theta \Upsilon' +
\gTTi \partial_\theta^2 \Xi \\
& + 3 \left(2 \sigma' + \lngTTp \right)
\Psi' - \left. \left(\frac{3}{2} \sigma' + \lngTTp
\right) \Xi' + \left[\frac{3}{2} \sigma' + 2 \left(\lngTTp\right)
\right] \Omega' - 3 \gTTi \sigma' \partial_\theta \Upsilon
\right\} + G_{\mu \nu}  \left(\Psi - \Xi \right),
\end{aligned}
\end{equation}
for the purely brane part, while the mixed ones read
\begin{equation}
\begin{aligned}
\delta G_{\mu r} & = \frac{1}{2} \partial_\mu \Bigg\{3 \Psi' + \Omega'
- \gTTi \partial_\theta \Upsilon - \left(\frac{3}{2} \sigma' +
\lngTTp \right) \Xi + \left(-\frac{1}{2} \sigma' + \lngTTp \right)
\Omega \Bigg\},\\
\delta G_{\mu \theta} & = \frac{1}{2} \partial_\mu
\Bigg\{\partial_\theta\left(3 \Psi + \Xi \right) - \Upsilon' -
\left(\sigma' + \lngTTp \right) \Upsilon \Bigg\},
\end{aligned}
\end{equation}
and the purely bulk components are
\begin{equation}
\begin{aligned}
\delta G_{rr} & = \frac{1}{2} \ee^{-\sigma} \Box\left(3 \Psi +
\Omega\right) - 2 \gTTi \partial_\theta^2 \Psi + 2 \gTTi \sigma'
\partial_\theta \Upsilon - \left[3 \sigma' + 2 \left(\lngTTp \right)
\right] \Psi' - \sigma' \Omega',\\
\delta G_{\theta \theta} & = \frac{1}{2} \gTT \ee^{-\sigma} \Box
\left(3 \Psi + \Xi \right) - 2 \gTT \Psi'' + \gTT \sigma'
\left(\Xi' - 5 \Psi' \right) + \frac{1}{2} \gTT \left(4 \sigma'' + 5
\sigma'^2 \right) \left( \Xi - \Omega \right),\\
\delta G_{r \theta} & = -\frac{1}{2} \ee^{-\sigma} \Box \Upsilon +
2 \partial_\theta \Psi' - \sigma' \partial_\theta \Xi + \left[\sigma'
  - 2 \left(\lngTTp\right) \right] \partial_\theta \Psi -\frac{1}{2}
\left(4 \sigma'' + 5 \sigma'^2 \right) \Upsilon.
\end{aligned}
\end{equation}
\end{widetext}
where $\Box$ is the brane d'Alembertian defined above
[Eq.~(\ref{nabla})].

\subsection{Stress-energy tensor}

In terms of the underlying fields, the stress-energy tensor stemming from
Eq.~(\ref{eq:tmunumatt}) reads
\begin{equation}
\begin{aligned}
T_{\si{AB}} & =\frac{1}{4} \left(\DD_{\si A} \Phi \right)^\dag
\left(\DD_{\si B}
\Phi \right) + \frac{1}{4} \left(\DD_{\si B} \Phi \right)^\dag
\left(\DD_{\si A} \Phi \right) \\
& - g^{\si{CD}} \F_{\si{AC}} \F_{\si{BD}} -
g_{\si{AB}} \mathcal{L}_{\matter},
\end{aligned}
\end{equation}
with $\mathcal{L}_\matter$ given by Eq.~(\ref{eq:lag}); to zeroth
order, this is given by
\begin{equation}
\mathcal{L}_\matter = -\frac{1}{2} \gTTi \frac{Q'^2}{q^2} -
\frac{1}{2} \varphi'^2 - \frac{1}{2} \gTTi \varphi^2 Q^2 - V(\varphi),
\end{equation}
and to first order in metric and field perturbations, using
Eq.~(\ref{eq:perthiggs}), this perturbed matter Lagrangian reads
\begin{equation}
\begin{aligned}
\delta \mathcal{L}_\matter & = \gTTi \frac{Q'}{q} \left(\GperT' -
\partial_\theta \GperR \right) + \frac{1}{2} \gTTi \frac{Q'^2}{q^2}
\left(\Xi + \Omega \right) \\ & + \frac{1}{2} \varphi'^2 \Xi +
\frac{1}{2} \gTTi \varphi^2 Q^2 \Omega + \gTTi q \varphi^2 Q \GperT \\
& - \left[ \varphi' \dfrac{\Chi' + \Chi'^\dag}{2} + \gTTi \varphi Q^2
\dfrac{\Chi+\Chi^\dag}{2} \right. \\
& + \left. \gTTi \varphi Q \partial_\theta
\dfrac{\Chi -\Chi^\dag}{2i} + \dfrac{\dd V}{\dd \varphi}
\dfrac{\Chi+\Chi^\dag}{2} \right].
\end{aligned}
\end{equation}
The purely brane part of the perturbed stress-energy tensor is therefore
given by
\begin{widetext}
\begin{equation}
\begin{aligned}
\delta T_{\mu \nu} & = \ee^{\sigma} \eta_{\mu \nu} \Bigg\{-\gTTi
\frac{Q'}{q} \left(\GperT' - \partial_\theta \GperR \right) +
\left[\varphi' \dfrac{\Chi'+\chi'^\dag}{2} + \gTTi \varphi Q^2
\dfrac{\Chi + \Chi^\dag}{2} + \gTTi \varphi Q \partial_\theta
\dfrac{\Chi-\Chi^\dag}{2i} + \dfrac{\dd V}{\dd\varphi}
\dfrac{\Chi+\Chi^\dag}{2} \right] \\ & - \left(\frac{1}{2} \gTTi
\frac{Q'^2}{q^2} + \frac{1}{2} \varphi'^2 \right) \Xi -
\left(\frac{1}{2} \gTTi \frac{Q'^2}{q^2} + \frac{1}{2} \gTTi \varphi^2
Q^2 \right) \Omega + \left[\frac{1}{2} \gTTi \frac{Q'^2}{q^2} +
\frac{1}{2} \varphi'^2 + \frac{1}{2} \gTTi \varphi^2 Q^2 + V(\varphi)
\right] \Psi \\ & - \gTTi q \varphi^2 Q \GperT \Bigg\},
\end{aligned}
\end{equation}
while the mixed components are
\begin{equation}
\begin{aligned}
\delta T_{\mu r} & = \partial_\mu \Bigg\{- \gTTi \frac{Q'}{q}
\left(\GperT - \partial_\theta \Gper \right) + \varphi' \dfrac{\Chi
+ \Chi^\dag}{2} \Bigg\},\\
\delta T_{\mu \theta} & =
\partial_\mu \Bigg\{ \frac{Q'}{q} \left(\GperR - \Gper' \right)
 +\varphi Q \dfrac{\Chi-\Chi^\dag}{2i} -q \varphi^2 Q \Gper
\Bigg\},
\end{aligned}
\end{equation}
and the bulk ones
\begin{equation}
\begin{aligned}
\delta T_{rr} & = -\gTTi \frac{Q'}{q}\left(\GperT' -
\partial_\theta \GperR \right) + \left[ \varphi' \dfrac{\Chi'+\Chi'^\dag}{2}
 - \gTTi \varphi Q^2 \dfrac{\Chi + \Chi^\dag}{2} - \gTTi \varphi Q
\partial_\theta \dfrac{\Chi-\Chi^\dag}{2i} - \dfrac{\dd V}{\dd \varphi}
\dfrac{\Chi+ \Chi^\dag}{2} \right] \\ & - \left[\frac{1}{2} \gTTi \varphi^2
Q^2 + V(\varphi) \right] \Xi - \left(\frac{1}{2} \gTTi
\frac{Q'^2}{q^2} - \frac{1}{2} \gTTi \varphi^2 Q^2 \right) \Omega +
\gTTi q \varphi^2 Q \GperT ,\\ 
\delta T_{\theta \theta} & =
-\frac{Q'}{q} \left(\GperT' - \partial_\theta \GperR \right) +
\left[-\gTT \varphi' \dfrac{\Chi'+\Chi'^\dag}{2} + \varphi Q^2
\dfrac{\Chi + \Chi^\dag}{2} + \varphi Q \partial_\theta \dfrac{\Chi -
\Chi^\dag}{2i} - \gTT \dfrac{\dd V}{\dd \varphi} \dfrac{\Chi + \Chi^\dag}{2}
\right] \\ & +
\left(\frac{1}{2} \gTT \varphi'^2 - \frac{1}{2}
\frac{Q'^2}{q^2}\right) \Xi - \gTT \left[ \frac{1}{2} \varphi'^2 +
V(\varphi) \right] \Omega - q \varphi^2 Q \GperT,\\ 
\delta T_{r \theta} & = \left[\varphi Q \dfrac{\Chi'-\Chi'^\dag}{2i} -
 \varphi'Q \dfrac{\Chi - \Chi^\dag}{2i} + \varphi' \partial_\theta
\dfrac{\Chi + \Chi^\dag}{2} \right] 
+ \left[\frac{1}{2} \gTTi \frac{Q'^2}{q^2} - \frac{1}{2} \varphi'^2
 - \frac{1}{2} \gTTi \varphi^2 Q^2 - V(\varphi) \right] \Upsilon 
- q \varphi^2 Q \GperR.
\end{aligned}
\end{equation}
\end{widetext}

\subsection{Faraday tensor}

In order to directly derive the perturbed Maxwell equations from
Eq.~(\ref{eq:max}), we have used the following perturbed Faraday
tensor whose purely brane components vanish
\begin{equation}
\delta \F_{\mu \nu} = 0 = \delta \F^{\mu \nu},
\end{equation}
and with the mixed parts
\begin{equation}
\begin{aligned}
\delta \F_{\mu r} & = \partial_\mu \left(\GperR - \Gper' \right),\\
\delta \F^{\mu r} & = - \ee^{-\sigma} \eta^{\mu \nu} \partial_\nu
\left(\GperR - \Gper' \right),\\ \delta \F_{\mu \theta} & =
\partial_\mu \left(\GperT - \partial_\theta \Gper \right),\\ \delta
\F^{\mu \theta} & = - \frac{\ee^{-(\sigma+\gamma)}}{\rdim^2} \eta^{\mu
\nu} \partial_\nu \left(\GperT - \partial_\theta \Gper \right).
\end{aligned}
\end{equation}
The only nonvanishing purely bulk components end up being
\begin{equation}
\begin{aligned}
\delta \F_{r \theta} & = \GperT' - \partial_\theta \GperR,\\ \delta
\F^{r \theta} & = \gTTi \left[\frac{Q'}{q} \left(\Xi + \Omega \right)
+ \GperT' - \partial_\theta \GperR \right].
\end{aligned}
\end{equation}
Owing to these formulas, one can calculate the perturbed Faraday
tensor divergence involved in Eq.~(\ref{eq:max}) by means of
\begin{equation}
\delta \F^{\si{AB}}_{\ \ \ ;\si{A}} = \partial_{\si A} \delta \F^{\si{AB}} +
\Gamma^{\si A}_{\ \si{DA}} \delta \F^{\si{DB}},
\end{equation}
where we have used the antisymmetry property of $\delta
\F_{\si{AB}}$. The perturbed left-hand side of Eq.~(\ref{eq:max}) can
be, in turn, expressed in terms of the perturbed matter fields by
means of Eqs.~(\ref{eq:d}) and (\ref{eq:perthiggs}) to give the
perturbed Maxwell equations (\ref{eq:pertmax_mu}) to
(\ref{eq:pertmax_theta}).

\section{The 6D cylindrical solution}\label{App:6dLambda}

In this appendix, we closely follow the analysis of Ref.~\cite{Linet:1984},
generalized to the six-dimensional case, and show that the only
gravity confining solution to Einstein equations in the presence of a
negative cosmological constant is the one used throughout this
article.

We start with the most general static, cylindrically symmetric line
element for a hyperstring inside which one assumes also rotation
invariance, namely
\begin{equation}
\dd s^2 =  g_\three(\rdim)\dd t^2-\dd\rdim^2 -g_\one (\rdim) \left(\dd
x^2 + \dd y^2 + \dd z^2 \right) -g_\two(\rdim)\dd\theta^2.
\label{eq:ds2gen}
\end{equation}
Note that if one also demands Lorentz invariance along the vortex
world sheet, then one would restrict attention to the subset of
solutions for which $g_\one (\rdim) = g_\three(\rdim)$, if it exists.

Setting $u^2\equiv g_\one^3 g_\two g_\three = -\det (g_\si{AB})$,
Einstein equations take the form
\begin{equation}
\left( \frac{u}{g_i} g'_i\right)' + \Lambda u =0,
\label{eq:gi1}
\end{equation}
where a prime indicates a derivative with respect to $\rdim$, and
\begin{equation}
3\frac{g'_\one g'_\two}{g_\one g_\two}+
3\frac{g'_\one g'_\three}{g_\one g_\three} +
\frac{g'_\two g'_\three}{g_\two g_\three} + 3\left(
\frac{g'_\one}{g_\one} \right)^2 + 4 \Lambda =0,
\label{eq:gi2}
\end{equation}
which can be combined to yield the simple equation for the metric
determinant, namely
\begin{equation}
u'' + \frac{5}{2} \Lambda u =0.
\label{eq:udet}
\end{equation}
With $\Lambda < 0$, the general solution of Eq.~(\ref{eq:udet}) is
\begin{equation}
u=u_+ \ee^{\rdim/\rdim_\Lambda} + u_- \ee^{-\rdim/\rdim_\Lambda}
\label{eq:solu}
\end{equation}
with $\rdim_\Lambda^{-2} = -5\Lambda/2$. We shall come back shortly to
this solution.

Eq.~(\ref{eq:udet}) can be integrated, with the result that
\begin{equation}
u^{\prime 2} = \frac{u^2}{\rdim^2_\Lambda} + K^2,
\label{eq:uint}
\end{equation}
where $K$ is an arbitrary real constant, \ie $K^2 >0$. This latter
requirement stems from the fact that, in order to have actual
cylindrical coordinates with $0\leq\theta\leq2\pi$ and to avoid a
singularity at the symmetry point $\rdim =0$, one must in general
impose that $\lim_{\rdim\to 0} u(\rdim)=0$.

Expanding and integrating Eqs.~(\ref{eq:gi1}) lead to the solution
\begin{equation}
\frac{g_i'}{g_i} = \frac{K K_i}{u} + \frac{2u'}{5 u},
\label{eq:giu}
\end{equation}
where the otherwise arbitrary constants $K_i$ are related through
$$
3K_\one + K_\two + K_\three = 0,
$$
coming from the definition of $u$ in terms of $g_i$, and
$$
2 K_\one (K_\two + K_\three) + K_\two K_\three = -\frac{8}{5},
$$
which is nothing but a rewriting of Eq.~(\ref{eq:gi2}).

Now, with $u(0)=0$, the solution (\ref{eq:solu}) or a direct
integration of (\ref{eq:uint}) gives
\begin{equation}
u = K \rdim_\Lambda \sinh\left(\frac{\rdim}{\rdim_\Lambda}\right),
\label{eq:usinh}
\end{equation}
and therefore the metric functions read
\begin{equation}
g_i(\rdim) = g_i^{(0)} \left[ \sinh
\left(\frac{\rdim}{\rdim_\Lambda}\right) \right]^{\frac{2}{5}} \left[
\tanh \left(\frac{\rdim}{2\rdim_\Lambda}\right) \right]^{K_i},
\label{eq:giKi}
\end{equation}
where the $g_i^{(0)}$ are three constants of integrations satisfying
$$
\left[g_\one^{(0)}\right]^3 g_\two^{(0)} g_\three^{(0)} = \frac{2
K^2}{5\Lambda}
$$ and $\left[ u'(0) \right]^2 = K^2$. Eq.~(\ref{eq:giKi}) is but the
6 dimensional generalization for nonvanishing cosmological constant
of the Kasner metric~\cite{Kasner:1925, Peter:prehistoric} already
obtained in Ref.~\cite{Linet:1984}.

For the purpose of confining gravity, one needs a space with finite
transverse volume $\mathcal{V}_{(2)}^\perp \equiv \int
\dd\theta\dd\rdim\sqrt{-g} = 2\pi\int u(\rdim)\dd\rdim$, \ie one
demands that $u(\rdim)$ goes asymptotically to zero faster than
$1/\rdim$. The solution (\ref{eq:usinh}) leads however to
$\mathcal{V}_{(2)}^\perp=2\pi K \cosh (\rdim/\rdim_\Lambda)
|_0^\infty$, which is unbounded unless $K\to 0$, in which case it is
undefined. The general solution is thus useless as one needs to impose
$K=0$. However, with $u_+=0$ in Eq.~(\ref{eq:solu}), the transverse
volume remains finite, so gravity ends up being actually confined on
the hyperstring core. In this case, since $u(0)\not=0$, the solution
is singular at $\rdim=0$, and is thus incomplete, being unable to
describe the vortex location itself. The warped metric
\begin{equation}
\dd s^2 = \ee^{-\rdim/\rdim_\lambda} \left( \eta_{\mu\nu}\dd x^\mu \dd
x^\nu - \dd\theta^2\right) -\dd\rdim^2
\label{eq:warp}
\end{equation}
is thus only asymptotically valid. Note that this is in fact not
problematic in the framework of the Abelian Higgs vortex since the
point at $\rdim=0$ cannot be described by the vacuum Einstein
equations.

An interesting point concerning the finite-volume metric
(\ref{eq:ds2gen}) is that, seen from far away from the vortex, the
warp factor can be interpreted as a missing angle, just like in the
simpler case of a Nambu-Goto string in Minkowski space. However, in
the case at hand, the requirements of both cylindrical and maximal
symmetries now demands a missing angle of $2\pi$, as can be seen by
evaluating the diameter of a circle $\rdim=$const., at a distance
$\rdim\to\infty$ and noting that this diameter vanishes with the warp
factor. 

One can also note that changing the radial coordinates into
$$
\rdim \to \bar\rdim = \bar\rdim_\Lambda
\exp\left(-\displaystyle\frac{\rdim}{2\rdim_\Lambda}\right) ,\qquad
\theta\to\bar\theta = \frac{\theta}{\bar\rdim_\lambda},
$$
with $\bar\rdim_\Lambda = 2 \rdim_\Lambda = \sqrt{10/|\Lambda|}$,
transforms the metric into
\begin{equation}
\dd s^2 = -\left(\frac{\bar\rdim}{\bar\rdim_\Lambda}\right)^2 \left(
\eta_{\mu\nu}\dd x^\mu \dd x^\nu\right) - \bar\rdim^2\dd\bar{\theta}^2
-\left(\frac{\bar\rdim_\lambda}{\bar\rdim}\right)^2 \dd\bar\rdim^2.
\label{eq:ds2R}
\end{equation}
With these new coordinates, the hyperstring is
located at $\bar\rdim=\bar\rdim_\Lambda$, and $\bar\theta$ is a compact
coordinate, varying between $0$ and $2\pi/\bar\rdim_\Lambda$.

Finally, it is worth pointing out that when a time dependence is
allowed for in either (\ref{eq:warp}) or (\ref{eq:ds2R}), and metric
variables are assumed separable, with $g_i (r)\to g_i(r) \Pi_i(t)$,
then one still has a valid solution provided $\Pi_\one=\Pi_\three = C
(t-t_0)^2 = \Pi_\two^{-1}$, with $C$ and $t_0$ arbitrary constants of
integrations. This solution, which is also of constant scalar
curvature $R=3\Lambda$, corresponds, to a shrinking extra dimension (a
circle located at a given coordinate distance from the vortex gets
smaller with time) and an expanding brane interior. As discussed
above, in the framework of the present article, this solution is only
asymptotically valid. Nevertheless, according to the background field
solutions (see Sect.~\ref{Sec:IV}), this shrinking might be associated
with a growing conical singularity in the vortex core, \ie with a
time-dependent $v_\zero$. It is a peculiarity of this number of
dimensions that the scale factor in the brane in this model evolves as
$\propto t^2$, which, recalling the time coordinate to be conformal
time, is the expected evolution of a matter-dominated universe.

\bibliography{bibadsstring}

\end{document}